\documentclass[aip,jcp,a4paper,reprint,onecolumn]{revtex4-1}

\usepackage[fleqn]{amsmath}
\usepackage{amssymb}
\usepackage[dvips]{graphicx}
\usepackage{color}
\usepackage{tabularx}
\usepackage{algorithm}
\usepackage{algorithmic}


\newcommand{\redc}[1]{{\color{black} #1}}

\newcommand{\vect}[1]{\textbf{\textit{#1}}}
\newcommand{\dd}[1]{\textsf{#1}}

\newcommand{\AT}{{\textrm{{AT}}}}

\newcommand{\CG}{{\textrm{CG}}}
\newcommand{\HY}{{\Delta}}
\newcommand{\rdf}{{\textrm{rdf}}}
\newcommand{\corr}{C^{(3)}}
\newcommand{\thf}{{\textrm{th}}}
\newcommand{\dof}{{\textrm{dof}}}

\newcommand{\ext}{{\textrm{extra}}}
\newcommand{\exc}{{\textrm{exc}}}
\newcommand{\thermo}{{\textrm{Q}}}
\newcommand{\hadress}{{\textrm{H}}}

\newcommand{\kin}{\textrm{kin}}

\begin{document}

\title{Grand-canonical-like molecular-dynamics simulations by using an adaptive-resolution technique}
\author{Han Wang}
\affiliation{Institute for Mathematics, Freie Universit\"at Berlin, Germany}
\author{Carsten Hartmann}
\affiliation{Institute for Mathematics, Freie Universit\"at Berlin, Germany}
\author{Christof Sch\"utte}
\affiliation{Institute for Mathematics, Freie Universit\"at Berlin, Germany}
\author{Luigi Delle Site}
\email{luigi.dellesite@fu-berlin.de}
\affiliation{Institute for Mathematics, Freie Universit\"at Berlin, Germany}

\begin{abstract}
  In this work, we provide a detailed theoretical analysis, supported by numerical tests, of the
  reliability of the adaptive resolution simulation (AdResS) technique in sampling the Grand Canonical
  ensemble. We demonstrate that the correct density and radial distribution
  functions in the hybrid region, where molecules change resolution, are two necessary conditions for
  considering the atomistic and coarse-grained regions in AdResS equivalent to subsystems
  of a full atomistic system with an accuracy up to the second order with respect to the probability distribution of the system. Moreover, we show that the
  work done by \redc{the thermostat and} a thermodynamic force in the transition region is formally equivalent to balance the chemical
  potential difference between the different resolutions. From these results follows the main conclusion that the atomistic region exchanges
  molecules with the coarse-grained region in a Grand Canonical
  fashion with an accuracy up to (at least) second order. Numerical tests, for the relevant case of liquid water at ambient conditions, are carried out to strengthen the conclusions of the theoretical analysis.
  Finally, in order to show the computational convenience of AdResS as a Grand Canonical set up, we compare our method to the Insertion Particle Method (IMP) in its most efficient computational implementation.
  This fruitful combination of theoretical principles and numerical evidence candidates the adaptive resolution technique as a natural, general and efficient protocol for Grand Canonical Molecular Dynamics for the case of large systems.
\end{abstract}

\maketitle

\section{Introduction}
The Adaptive Resolution Simulation (AdResS) scheme~\cite{jcp,pre} is a
method that in a concurrent fashion simulates a molecular system treated with different
resolutions in different regions of space.  The word ``resolution''
here refers to the amount of details included in a molecular model: the higher 
resolution corresponds to a more precise model, but at higher computational costs. The advantage is obvious, it
keeps track of both: the local fine-grained processes in the
high-resolution regions and the large scale behavior, requiring a less demanding computational effort compared to the system that, as a
whole, would be described by the high-resolution model. The key feature of
AdResS is that it allows for an on-the-fly change of resolution when a
molecule travels from a high-resolution region to a low-resolution
region and vice versa. Moreover recent research~\cite{prlgc, rdfcorr}
has numerically demonstrated that different regions (with different resolutions) reach a
thermodynamic equilibrium as if the whole system were equilibrated
under the high-resolution description. Therefore one is tempted to state that a region with a
certain resolution exchanges molecules with the rest of the system in
a Grand Canonical fashion. This work
studies the minimal necessary conditions, such that AdResS
performs effective Grand Canonical simulations.  
Moreover we shed light on the level accuracy of
the sampling in a Grand Canonical fashion; it will be shown that indeed the sampling can be considered 
of the Grand Canonical type if we accept an accuracy on the probability distribution of the system up to the second order.
Though the second order is numerically satisfying, higher level of accuracy can be systematically reached by requiring to match, in the transition region, higher orders of the probability distribution.

From the numerical point of view, for systems which are large enough, a (relatively) small subsystem of a full atomistic region is a 
natural Grand Canonical ensemble, for this reason the approach we use here is that of showing the equivalence between the atomistic region of AdResS and the same region in a full atomistic simulation.
Extensive numerical tests, for the relevant case of liquid water at ambient conditions, are presented in order to show that the hypothesis done within the theoretical analysis are justified from the numerical point of view. Beyond our expectations, we have found that in the atomistic region of AdResS not only the accuracy can be tuned up to the second order, but, without any additional correction, even the three body correlation function of molecular centers of mass (COM) turns out to be the same as that of a full atomistic simulation. Finally, a further numerical test was carried out: the coarse-grained molecules are substituted by a liquid of spheres interacting via the Weeks-Chandler-Andersen (WCA)  potential \cite{wca} which has no reference to the atomistic resolution. We show that also in this case in the atomistic region, due to the work of the filter of the transition region, an accuracy up to the third order in the probability distribution is achieved.
The results of this work allow us to make a step further in the possibility of employing the AdResS scheme as a natural, general numerical protocol for truly Grand Canonical Molecular Dynamics simulations
independently from the nature of the particles acting as a reservoir in the coarse-grained region. The word {\it ``natural''} refers to the fact  that strictly speaking in statistical mechanics the Grand Canonical ensemble is defined in operative terms as a subsystem of an infinitely large system. Of course in simulation systems are never infinitely large, however they can be large enough to numerically satisfy this condition. 
In this sense, in our approach molecules are exchanged between the atomistic domain and the coarse-grained reservoir in a straightforward (i.e.{\it ``natural''}) dynamical way. The advantage compared to previous Grand Canonical schemes for molecular dynamics is that the proper exchange of particles {requires neither the knowledge or the calculation of the chemical potential, nor additional expensive insertion or removing of particles \cite{pet1,pet2,pet3,pet4,pet5,flo}; we actually compare the computational costs of our method to those of the IPM and show that for dense liquid systems AdResS is more efficient.} 
The paper is organized as follows: first we briefly describe the essential principles on which AdResS is based, then we discuss the approximations under which one can write the probability distribution of the system in terms of Grand Canonical partition function. Next we show that \redc{the work of the thermostat plus that of} the thermodynamic force balances the difference in chemical potential; this idea allows us to derive the necessary conditions of simulation to reach the accuracy in terms of orders of the probability distribution of the system. 
Finally the numerical results {and the efficiency of the method} are discussed.

\begin{figure}
  \centering
  \includegraphics[width=0.9\textwidth]{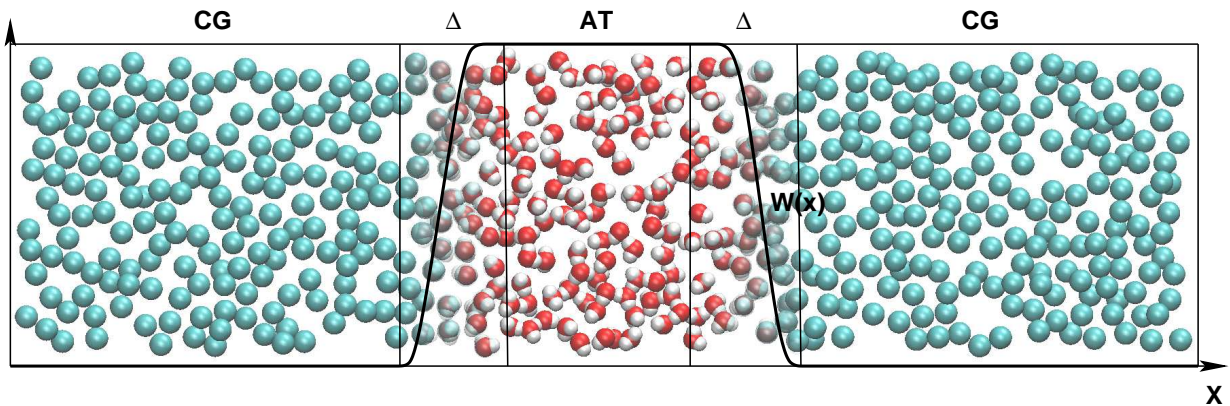}
  \caption{A schematic representation of liquid water in AdResS. CG indicates the region at coarse-grained resolution, AT that at atomistic resolution, and finally $\Delta$ the transition region where molecules change resolution according to a smooth switching function $w(x)$.}
  \label{fig:adress-water}
\end{figure}

\section{A brief description of the AdResS scheme}

In this work, the higher resolution refers to the atomistic
description (AT) of a molecule, while the lower resolution refers to
its corresponding coarse-grained (CG) model (for a pictorial representation of a typical AdResS set up, see Fig.\ref{fig:adress-water} for the case of liquid water).
We assume that the dynamics of the
system is subject to the Langevin equation, which is the thermostat
usually used in the AdResS simulations.
We denote the
phase space variable $\vect x = (\vect r_1,\cdots,\vect r_N, \vect v_1,\cdots,\vect v_N)$; $\vect r_i,
\vect v_i$ are the center-of-mass
degrees of freedoms (DOFs) of the molecules.
For simplicity we do not explicitly write the
  atomistic DOFs, but one should keep in mind that the interaction
  between two  molecules with atomistic resolution
  are given by the sum of all pair-wise atomic interactions.
  Also for the sake of simplicity,
  we do not explicitly distinguish between the COM and the atomistic coordinates,
  using a generalized formalism where the coordinates are indicated simply as $\vect r_i$, however the interpretation of such a notation is
  made clear by the specific context.
If the
system is conservative, 
then one obtains the equilibrium density distribution $p(\vect
x)\propto e^{-\beta\mathcal H(\vect x)}$, where $\beta = 1/k_BT$ is the inverse temperature.
  In this paper, we assume that
  the kinetic part of the Hamiltonian is decoupled from the configurational part, then
  it is usually more convenient to consider the configurational
  probability distribution, namely $ p(\vect r_1,\cdots,\vect r_N) =
  \int p(\vect r_1,\cdots,\vect r_N, \vect v_1,\cdots,\vect v_N)\,\dd d\vect v_1,\cdots,\dd d\vect v_N$.
Throughout the paper, when we mention the $i$-th order of a
  multi-body configurational probability distribution, e.g.
  $p(\vect r_1, \cdots, \vect r_N)$, we mean its $i$-th marginal
  distribution, which is defined by
  $p^{(i)}(\vect r_1, \cdots, \vect r_i)
  = \int p(\vect r_1, \cdots, \vect r_i, \vect r_{i+1}, \cdots, \vect r_N)\,
  \dd d \vect r_{i+1}\cdots\dd d \vect r_N$.
  It is obvious that
  whenever we have the $i$-th order accuracy of a probability distribution,
  any lower  order is automatically accurate.
In the AdResS scheme,
different resolutions in the system are described by a weighting
function $w(\vect r)$. Usually the higher resolutions is
denoted by $w = 1$, while the lower resolution is denoted by $w = 0$.
Between the higher and lower resolutions, a hybrid region allows a 
molecule to have both (interpolated) resolutions. The weighting function changes smoothly
from 0 to 1; one possible form of such a function is:
\begin{align}\label{eqn:new-w}
  w(\vect r) =
  \left\{
    \begin{array}{lcl}
      1 &\quad& \chi(\vect r) < 0\\
      1  && 0 < \chi(\vect r) < r_c\\
      \cos^2\big[\frac{\pi}{2(d_{\HY} - r_c)} (\chi(\vect r) - r_c)\big] && r_c < \chi(\vect r) < d_{\HY} \\
      0 &&  d_{\HY}  < \chi(\vect r).
    \end{array}
  \right.
\end{align}
Where $\chi(\vect r)$ is the distance between the
molecule at $\vect r$ and the boundary of the higher resolution
region, $\chi(\vect r) < 0$ means that the molecule is in the high resolution
region, $d_{\HY}$ is the thickness of the hybrid region, $r_c$ is the
cut-off radius of the atomistic interactions.  The intermolecular force is modeled by the following
interpolation formula \cite{rdfcorr}:
\begin{align}
  \vect F_{ij} =
  w_i w_j\vect F^{\AT}_{ij} +
  (1-w_i w_j)\vect F^{\CG}_{i j} +
  w_i w_j (1-w_i w_j)\vect F_{ij}^{\rdf}.
\label{grf}
\end{align}
Where $w_i = w(\vect r_i)$ and $w_j = w(\vect r_j)$.
$ \vect F^{\AT}_{ij}$, $ \vect F^{\CG}_{ij}$
are the intermolecular interaction of the atomistic and coarse-grained
resolutions, respectively.
$\vect F_{ij}^{\rdf}$ is a force that corrects the COM-COM radial distribution function (RDF) in the transition region $\Delta$ ~\cite{rdfcorr}. A further one body force, named thermodynamic force, ${\vect F}_\thf$, is applied to ensure the correct thermodynamic equilibrium of the system (for specific details, see Ref.~\onlinecite{prlgc}):
\begin{equation}
  {\vect F}_{i}=
  \sum_{j}{\vect F}_{ij}+
  {\vect F}_\thf(\vect r_i),
\label{mody}
\end{equation}
which is defined by
\begin{equation}
  p_{\AT}+
  \rho_0
  \int_{\Delta} {\vect F}_\thf(\vect r)\,\dd d\vect r
  =p_{\CG},
  \label{thf}
\end{equation}
where $p_{\AT}$ is the pressure of the atomistic resolution, $p_{\CG}$ that of the coarse-grained resolution, $\rho_{0}$ is the equilibrium number density, corresponding to that of a full atomistic simulation.
Such a thermodynamic force has been derived by empirical considerations regarding the equilibrium of open systems and is based on {\it forcing} the equality of the grand potential of the atomistic part with the rest of the system.
This is then reduced to provide a balancing force to the difference of pressure at the wished uniform density of equilibrium. Regarding this point, one of the main results of this paper is that the \redc{thermodynamic force and the thermostat perform a work in the transition region which balances the difference in chemical potential} between the different regions. As a consequence (equilibrium of the chemical potential) the interpretation of the AdResS simulation as a Grand Canonical sampling is enforced even further.
 A crucial point of this work is the following: the 
AdResS scheme is not Hamiltonian~\cite{presolo,prlcomm}; however, under the hypothesis of fixing DOFs in the hybrid region for a statistical analysis, the atomistic and
coarse-grained regions can be considered (in good approximation) Hamiltonian; this line of thought gives rise to the basic
idea of the present study as it is presented in the next sections. Moreover Refs.\cite{presolo,prlcomm} also show that a space dependent interpolation of potentials, which would lead to a direct Hamiltonian approach, is neither satisfactory from the point of view of the physical consistency, nor for the computational efficiency in improving in a systematic way the accuracy of the results in the $AT$ region; instead this latter is one of the main results of this work.

\section{Theoretical considerations}
\subsection{The outline of the basic idea}
Here we denote the degrees of freedoms and number of particles in the
atomistic region (AT), hybrid region ($\HY$) and the coarse-grained
region (CG) by $(\vect x_1, N_1)$, $(\vect x_2, N_2)$ and $(\vect x_3,
N_3)$, respectively. Therefore, the target is to prove that the atomistic
region is subject to the Grand Canonical statistics: 
\begin{align}\label{eqn:p}
  p(\vect x_1, N_1) = \frac{1}{\mathcal Q_1}
  e^{\beta\mu_{\AT} N_1 - \beta \mathcal H_{N_1}^{\AT}(\vect x_1)} 
\end{align}
where the partition function $\mathcal Q_1$ is defined by
\begin{align}
  \mathcal Q_1 =
  \sum_{N_1}\int
  \dd d\vect x_1\,
  e^{\beta\mu_{\AT} N_1 - \beta \mathcal H_{N_1}^{\AT}(\vect x_1)}
\end{align}
The marginal probability of finding $N_1$ molecules in the
AT region is:
\begin{align}\label{eqn:p-2}
  p(N_1) = \int\dd d\vect x_1\, p(\vect x_1, N_1)
  =
  \frac{1}{\mathcal Q_1}e^{\beta\mu_{AT} N_1}Q_{N_1}
\end{align}
where $Q_{N_1}$ is the partition function for a canonical ensemble
with $N_1$ atomistic molecules:
\begin{align}
  Q_{N_1}  =
  \int\dd d\vect x_1\,
  e^{ - \beta \mathcal H_{N_1}^{\AT}(\vect x_1)}
\end{align}
Let us consider
$p(\vect x_1, N_1) = p(\vect x_1 | N_1)\,p(N_1)$. Then,
from Eq.~\eqref{eqn:p} and \eqref{eqn:p-2},
the conditional probability $p(\vect x_1 | N_1)$
for a truly Grand Canonical ensemble turns out to be
\begin{align}\label{eqn:p-1}
  p(\vect x_1 | N_1) &= \frac{1}{Q_{N_1}} e^{-\beta \mathcal H_{N_1}^{AT}(\vect x_1)} \end{align}
The key point of our argumentation is the following: we want to compare
the distributions (in the various regions) of the AdResS simulation
with the corresponding ones of a full atomistic reference system, which is ideally divided
in subregions corresponding to the AT, $\HY$ and CG regions of the AdResS
set up. The first step in our procedure is to fix the number of molecules in the atomistic
region and consider the conditional probability $p(\vect x_1 |
N_1)$. If the AdResS set up would sample the space in a Grand Canonical fashion, then this
probability should be the same as the corresponding one of a full
atomistic reference system, namely Eq.~\eqref{eqn:p-1}.  Then, if the
probability of finding $N_1$ molecules in the atomistic region is also
the same as the full atomistic reference system, namely
Eq.~\eqref{eqn:p-2}, we can safely state that the atomistic region of AdResS
samples configurations in a Grand Canonical fashion. In fact as underlined before, it must be noticed that a subsystem of full atomistic system is a natural Grand Canonical ensemble in the thermodynamic limit (see, e.g., Ref.~\onlinecite{lanford1973entropy}). 
The thermodynamic limit in our case is intended in such a way that at a fixed density, the size of
  both the AT and CG regions tend to infinite, and, at the same time,
  the ratio between the extension of AT and that of the CG regions goes to zero; as underlined before, in practical terms, in numerical simulation the system size does not go to infinity, however it can be large enough so that within a certain numerical accuracy the hypothesis of thermodynamic limit holds.

\subsection{The conditional probability density $p(\vect x_1 | N_1)$}  \label{sec:p-1}

To prove the above statement about $p(\vect x_1 | N_1)$, we divide it into two parts:
\begin{align}\label{eqn:divide-cond-p}
  p(\vect x_1 | N_1) = \sum_{N_2}\int
  p(\vect x_1 | N_1; \vect x_2, N_2) \,
  p(\vect x_2, N_2 | N_1)
  \,\dd d\vect x_2
\end{align}
where $p(\vect x_1 | N_1; \vect x_2, N_2)$ is the probability density obtained by fixing the
coordinates and number of particles in the region $\HY$ and considering
the distribution of the DOFs in the AT region (see further clarifications below), while $p(\vect x_2, N_2 | N_1)$ is the distribution in the hybrid region, conditional on the number $N_{1}$ of particles in the atomistic region. We now comment on the underlying assumptions involved in the calculation of the conditional probability density: 

\begin{enumerate}

\item If we use the weighting function \eqref{eqn:new-w},
then the AT region is interacting with the $\HY$ region
as if the $\HY$ region were part of the AT region because
all hybrid molecules interacting with the AT molecules have unit weight.

\item We assume that $p(\vect x_1 | N_1; \vect x_2, N_2)$  can be approximated by
\begin{align}\label{eqn:p-1-1}
  p(\vect x_1 | N_1; \vect x_2, N_2)
  \propto &\,
  e^{-\beta\mathcal H_{N_1}^{\AT}(\vect x_1; \vect x_2, N_2)}
\end{align}
where the Hamiltonian governing the physics of the molecules in the atomistic region is defined as:
\begin{align}\label{eqn:p-1-2}
  \mathcal H_{N_1}^{{\AT}}(\vect x_1; \vect x_2, N_2) = &\,
  \sum_{j=1}^{N_1}\frac12 m_i\vect v_i^2 + 
  \sum_{i,j=1}^{N_1}\frac12 U^{{\AT}}(\vect r_i - \vect r_j)  +
  \sum_{i=1}^{N_1}\sum_{j=N_1+1}^{N_2} U^{{\AT}}(\vect r_i - \vect r_j)   
\end{align}
which is exactly the same as for the equivalent subregion of the full atomistic system of reference. (The approximation is essentially based on the assumption that AT and $\Delta$ regions are only short-range correlated; this is discussed in the following point.)

\item We further assume that \emph{all} interactions in the system have finite range with given cut-off radius; the
  electrostatic interaction is treated by the reaction field method. Specifically, we suppose that 
  the AT region does not interact in a direct way with the CG region.
  For the case of liquid water at ambient conditions, it is reasonable to assume that the system is only short-range correlated, i.e. the AT region is only correlated with the $\HY$ region. We will show numerically that this is indeed the case.)
  We suppose that all interactions satisfy the superstability condition of Ref.~\onlinecite{ruelle1970}.
  
\item The hypothesis of fixed molecules in the $\HY$ region, must not be intended in dynamical terms. This means that during the dynamical evolution of the system one should not suppose the molecules in the $\HY$ region to be frozen in their positions while molecules in other regions are moving. Instead this hypothesis must be intended in the sense of statistical analysis; within a statistical framework it is intended that for {\bf a given configuration} of molecules in the $\HY$ region, the AT region explores a large (statistical) number of configurations. In such a case, the practical statistical analysis consists of taking a given configuration in the $\HY$ region, then consider the whole trajectory of a simulation and sort out all the configurations in the AT region corresponding to the given configuration in the $\HY$ region. If one repeats the process for a large number of fixed configurations of the $\HY$ region, then the data of the trajectory sorted according to this criterion would lead to a statistics for the AT region equivalent to that obtained by a sampling performed according to the {\bf given Hamiltonian} of the AT region (Eq.\ref{eqn:p-1-2}).

\end{enumerate}

In accordance with the third assumption, we ignore correlations between the molecules of $\HY$ and CG regions, in fact we even consider the DOFs in the $\HY$ region fixed 
  while sampling configurations in the AT and CG regions. 
The question then is, whether the probability $p(\vect
x_2, N_2 | N_1)$ in the $\HY$ region  is the same as that of the equivalent region in a fully atomistic reference
system. In general this will not be the case, so we 
state necessary conditions to enforce such an equivalence to lowest order in the configurational probability of the system; the necessary conditions are~\cite{rdfcorr}:
\begin{align}\label{eqn:p-1-cond1}
  \rho_{\HY}(\vect r) &= \rho_{\AT}(\vect r)\\\label{eqn:p-1-cond2}
  g_{\HY}( r) &= g_{\AT}(r)
\end{align}
The first order marginal distribution is the particle (or molecular) density 
$\rho_{\HY}(\vect r)$, while the second order marginal distribution is
the radial distribution function (RDF), $g_{\HY}(r)$. 
The necessary
conditions of the correct distribution $p(\vect x_1|N_1)$
are that these two distributions should be the same as those of the fully atomistic reference system. 
These are the minimal necessary conditions involving the basic DOF's, that is the molecular center of mass coordinates, however one may require a more specific accuracy by imposing that also any atom-atom $g(r)$ is matched to the corresponding function of the atomistic reference system. 
In general, the conditions of Eq.\ref{eqn:p-1-cond1} and Eq.\ref{eqn:p-1-cond2} would assure that at least at the first and second order $p(\vect x_2, N_2 | N_1)$ in the AdResS is the same as that of a full atomistic simulation. One needs to go at least at the second order, so that at the interface between the atomistic and transition region, the radial distribution functions of the atomistic part are not affected by artifact due to the deviation of the RDF's of the $\HY$ region from the correct atomistic reference. 
We have shown numerically how the RDF correction, can be numerically implemented~\cite{rdfcorr}.
Higher accuracy can be then systematically reached by enforcing the equivalence of higher orders of the distribution in the transition region.
Next, we must show that $p(N_{1})$ is the same in AdResS and in the reference full atomistic simulation. The basic outline of arguments will be given in a following section while the (long) explicit calculations are reported in the Appendix~\ref{app:1}, however, before proceeding further we must first treat a key ingredient of this equivalence, that is the thermodynamic force. This force, in fact, \redc{together with the thermostat}, is the crucial tool for assuring the thermodynamic and statistical equilibrium of the AdResS system when compared to the reference full atomistic simulation. In the next section, we will show how this force, derived on intuitive ground to enforce the equality of some basic thermodynamic relations, as shown in Eq.\ref{thf}, is, \redc{together with the action of the thermostat, the key ingredient for the balance of chemical potential between the various region at the desired density of equilibrium}. The balance of chemical potential is implicitly a strong argument in favor of the idea of AdResS as Grand Canonical-like scheme (i.e. molecules are exchanged between the AT and CG region in conditions of equilibrium).

\subsection{Thermodynamic force: from empirical intuition to strict formalization within the Grand Canonical framework}
Intuitive thermodynamic considerations led to formula \eqref{thf}. In this section we show that, despite the argument used in Eq.\ref{thf} from Refs.\cite{prlgc,rdfcorr} has a strong empirical component, one can formally justify this force
as a tool to balance the chemical potential of the various resolutions and, as a consequence,  a key aspect in interpreting AdResS as an effective Grand Canonical set up. 
\begin{figure}
  \centering
  \begin{minipage}[t]{0.49\linewidth}
  \includegraphics[width=0.6\textwidth]{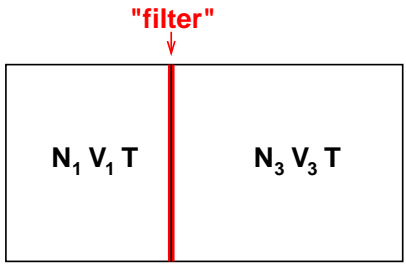}    
  \end{minipage}
  \caption{Schematic plot of the AdResS system in thermodynamic equilibrium. The thickness of the filter corresponds to $d_{\Delta}$.}
  \label{fig:tmp1}
\end{figure}
Here, two assumptions are made:
\begin{enumerate}
\item $N_2\ {\ll}\ N_1 \ll N_3$: the second inequality corresponds to
  the thermodynamic limit of the Grand Canonical ensemble. The first
  one actually assumes that the $\HY$ region is infinitely thin so that it
  can be viewed as an infinitesimal membrane that allows free exchange of molecules from
  the AT region to the CG region and vice versa.
\item We have 
  \begin{equation}
    \mathcal H(\vect x_1, N_1; \vect x_3, N_3) =
    \mathcal H_{N_1}^{\AT}(\vect x_1) + \mathcal H_{N_3}^{\CG}(\vect x_3). 
  \end{equation}
  The latter is reasonable if the system is
  short-range correlated.
  
\end{enumerate}
The equilibrium in the AT and CG region is assured by the membrane $\HY$ via the action of thermodynamic force and of \redc{the thermostat}. Conceptually, we assume there is an infinitely thin
``filter'' (see Fig.~\ref{fig:tmp1}) located at the interface between
the AT and CG regions. When a molecule enters into
the AT region or leaves it, the filter does some work per molecule, $\omega_0$, on the atomistic system in order to assure the thermodynamic equilibrium.
Therefore, we add an empirical term in the Hamiltonian of the system
$N_1\omega_0$, related to the work done  by the filter to obtain configurations of the atomistic region with $N_{1}$ molecules. 
Having fixed the number of molecules, that is ideally considering case by case situations at fixed $N_{1},N_{3}$, and by following the
arguments of the last section, both the AT and the CG regions are
subject to the Boltzmann distribution.
Thus, the fixed-number partition function ($N_1$ in the AT region and $N_3$ in the CG region) of the system reads
\begin{align}\nonumber
  q(N,V,T)
  &= \frac1{N!}\int
  \dd d\vect x
  e^{-\beta
    [\mathcal H(\vect x_1, N_1; \vect x_3, N_3) +
    N_1\omega_0]}\\\nonumber
  &= \frac1{N!}\int
  \dd d\vect x_1\dd d\vect x_3\,
  e^{-\beta
    [\mathcal H_{N_1}^{\AT}(\vect x_1) +
    \mathcal H_{N_3}^{\CG}(\vect x_3) +
    N_1\omega_0]}\\\nonumber
  & = \frac{N_1!N_3!}{N!}
  e^{-\beta N_1\omega_0}
  \frac{1}{N_1!}\int\dd d\vect x_1 e^{-\beta\mathcal H_{N_1}^{\AT}(\vect x_1)}
  \frac{1}{N_3!}\int\dd d\vect x_3 e^{-\beta\mathcal H_{N_3}^{\CG}(\vect x_3)}\\
  & = \frac{N_1!N_3!}{N!}
  e^{-\beta N_1\omega_0}
  Q_{\AT}(N_1, V_1, T)\,
  Q_{\CG}(N_3, V_3, T) 
\end{align}
Considering the permutations of particles, the number of possibilities of
$N_1$ molecules being in the atomistic region and $N_3$ molecules being
in the coarse-grained region is  $\frac{N!}{N_1!N_3!}$.
Therefore, the partition function of the whole system reads
\begin{align}\nonumber
  Q(N,V,T) &= \sum_{N_1=1}^N
  \frac{N!}{N_1!N_3!} \frac{N_1!N_3!}{N!}
  e^{-\beta N_1\omega_0}
  Q_{\AT}(N_1, V_1, T)\,
  Q_{\CG}(N_3, V_3, T) \\
  &= \sum_{N_1=1}^N
  e^{-\beta N_1\omega_0}
  Q_{\AT}(N_1, V_1, T)\,
  Q_{\CG}(N_3, V_3, T) 
\end{align}
with natural relations:
\begin{align}
  N &= N_1 + N_3\\
  V &= V_1 + V_3
\end{align}
For convenience, we denote
\begin{align}
  \tilde q_{N_1} = 
  e^{-\beta N_1\omega_0}
  Q_{\AT}(N_1, V_1, T)\,
  Q_{\CG}(N - N_1, V_3, T).
\end{align}
Further let $\bar N_1$ be the value at which $\tilde q_{N_1}$ reaches
its unique maximum, namely $\tilde q_{\bar N_1} = \max \tilde q_{N_1}$. (The existence of a maximum follows from the superstability of the interactions that guarantees that particles do not cluster as $N\to\infty$; we further assume that it is unique.) Since $\tilde q_{N_1}$ is positive definite, a basic
observation is that
\begin{align}
  \tilde q_{\bar N_1}
  \leq
  \sum_{N_1=1}^N \tilde q_{N_1}
  \leq
  N \tilde q_{\bar N_1}. 
\end{align}
Due to the monotonicity of the logarithm, we have
\begin{align}
  \ln\tilde q_{\bar N_1}
  \leq
  \ln\sum_{N_1=1}^N \tilde q_{N_1}
  \leq
  \ln (N \tilde q_{\bar N_1})
  =
  \ln N + \ln\tilde q_{\bar N_1}.
\end{align}
If $\ln N \ll \ln \tilde q_{\bar N_1}$, Laplace's method yields (see, e.g., Ref.~\onlinecite[Sec.~4.3]{dembo1998}) 
\begin{align}
  \ln\sum_{N_1=1}^N \tilde q_{N_1}
  \approx
  \ln\tilde q_{\bar N_1},
\end{align}
(The validity of the antecedent will be discussed later.) Hence
\begin{align}
  \ln Q(N, V, T)
  \approx
  -\beta \bar N_1\omega_0 + 
  \ln Q_{\AT}(\bar N_1, V_1, T) + \ln Q_{\CG}(N - \bar N_1, V_3, T)
\end{align}
or equivalently
\begin{align}\label{eqn:a-energy-1}
  A(N, V, T)
  \approx
  \bar N_1\omega_0 +
  A_{\AT}(\bar N_1, V_1, T) + A_{\CG}(N - \bar N_1, V_3, T)
\end{align}
Where $A$ denotes the Helmholtz free energy. 
At this point, the crucial question is if the condition $\ln N \ll \ln \tilde q_{\bar N_1}$
holds, or equivalently
\begin{align}\label{eqn:condition}
  \ln N 
  \ll
  -\beta \bar N_1\omega_0 +
  \ln Q_{\AT}(\bar N_1, V_1, T) + \ln Q_{\CG}(N - \bar N_1, V_3, T)
\end{align}
Generally this is true: $\ln Q_{\CG}(N - \bar N_1, V_3, T)$
is proportional to the free energy $A_{\CG}(N - \bar N_1, V_3, T)$,
which is an extensive thermodynamic variable, so
$A_{\CG}(N - \bar N_1, V_3, T)$ is proportional to $N-\bar N_1$.
Due to the thermodynamic limit $N \gg \bar N_1$,
$A_{\CG}(N - \bar N_1, V_3, T)$ is proportional $N$, which
is much larger than $\ln N$ for $N\gg 1$; this
validates condition~\eqref{eqn:condition}.
We will see later (from Eq.~\eqref{eqn:pn1})
that the maximum $\bar N_1$ corresponds to the maximum
value of probability $p(N_1)$; this is also the average molecule
number in the AT region that is of statistical importance
under the thermodynamic limit.
As the right hand side of
Eq.~\eqref{eqn:a-energy-1} attains its maximum at $\bar
N_1$ when the other thermodynamic variables are kept fixed, differentiating it with respect to $N_1$ entails
\begin{align}
  \omega_0 = \mu_{\CG}(N - \bar N_1, V_3, T)  - \mu_{\AT}(\bar N_1, V_1, T)
\end{align}
This means that the difference in the chemical potential between the AT and CG
regions is taken care by the work of the filter.
Now, if the filter ensures an
equilibrium that is the same as the full atomistic reference system, then:
\begin{align}\label{eqn:w-mu-diff}
  \omega_0 = \mu_{\CG}(\rho_0V_3, V_3, T) - \mu_{\AT}(\rho_0 V_1, V_1, T)
\end{align}
$\rho_0$ is the number density at which the atomistic and coarse-grained
resolutions should match,
namely $\bar N_1 = \rho_0V_1$ and $N - \bar N_1 = \rho_0(V - V_1)$ should apply.
Eq.~\eqref{eqn:w-mu-diff} is a necessary
condition \redc{for the work required to the filter in order to have}
the correct equilibrium of the AT and CG regions.\\

\redc{As anticipated in the previous sections, the work of the filter has two components, one corresponding to the work of the thermodynamic force and the other corresponding to the work of the thermostat, i.e. $\omega_0 = \omega_\thf + \omega_\thermo$. We indicate with $\omega_\thf$ the work of thermodynamic force and with $\omega_\thermo$ that of thermostat in the transition region. The existence of $\omega_\thermo$ has been numerically proven in Ref.\cite{simon} and can be decomposed in two parts: $\omega_\thermo=\omega_\dof+\omega_\thermo^\ext$. $\omega_\dof$ is related to the thermalization of the degrees of freedom (rotational and/or vibrational) which are reintroduced/removed, and according to the equipartition theorem, one has $\frac{1}{2}k_BT$ per degree of freedom, while $\omega_\thermo^\ext$ is due to the absence of the energy conservation that is related to the different intermolecular interactions in the different regions of the system as shown in Ref.\cite{prlcomm}. In the Appendix~\ref{app:4} we will show how this term can be numerically calculated within a standard AdResS, thus allowing the explicit calculation of the chemical potential of a system.}

We remind the reader that the argument in this section is essentially based on a large deviations principle (namely, Laplace's method) for the number of particles in the AT region that entailed that $\bar N_{1}$ rather than $N_{1}$ can be considered as representative of the configuration realizations in the AT region. This implies that the particle fluctuations, $\Delta N_{1}$,
are negligible compared to $\bar N_{1}$ in the AT region. This hypothesis is valid if the AT region is large enough so that $\bar N_{1}$ is large. This point seems to be true in all numerical experiments done so far (see, e.g., Ref.~\onlinecite{debash}). In this way, we have shown the formal derivation of the \redc{thermodynamic force and the thermostat as tools} to balance the chemical potential between the two regions. For a Grand Canonical-like set up, the balance of chemical potential between open regions is a necessary condition to have the exchange of molecules in thermodynamic equilibrium and as a consequence we have formally justified why Eq.\ref{thf} is a necessary condition for AdResS to be considered an effective Grand Canonical set up.

\subsection{The number probability $p(N_1)$}
\label{sec:pn1}
In this section, we provide the basic arguments by which one can define at which level of approximation $p(N_{1})$
in AdResS is the same as in a full atomistic simulation. This is an important point
in order to justify the reliability of AdResS as a Grand
Canonical set up in terms of probability distribution. In fact if
$p(N_{1})$ in AdResS is very different from the corresponding one of a
full atomistic simulation, then clearly the AdResS method cannot be
considered valid since the artifacts due to the
different $p(N_{1})$ would give a non realistic description of the
system.
Essentially the arguments of this equivalence are two; the 1st order accuracy of  $p(N_{1})$, requires the balance of the chemical potential:
\begin{align}\label{eqn:mu-eq}
  \mu_{\CG} &= \mu_{\AT} + \omega_0
\end{align}
This, in the previous section, has been shown to hold because of the action of the \redc{thermodynamic force and the thermostat} in the thermodynamic limit. 
Whether or not the size of standard and feasible molecular simulations can be considered, effectively, in the thermodynamic limit will be checked later on with numerical tests; however we can anticipate that the answer is positive for systems whose size and time of simulation are nowadays routinely done.
The 2nd order of accuracy, requires the equality for the compressibility:
\begin{align}\label{eqn:kappa-eq}
  \kappa_{\CG} &= \kappa_{\AT}.
\end{align}
Eq.\ref{eqn:kappa-eq} implies that the COM-COM RDFs (here, again, COM stands for the center of mass, and RDF stands for the radial distribution function) of the atomistic and coarse-grained region are matched \cite{han}.
Essentially these are the physical requirements to show that up to the second order $p(N_1)$ in AdResS is the same as in the equivalent subregion of the full atomistic system of reference.
Because of the lengthy arguments, specific details are reported in Appendix~\ref{app:1}.

\section{A numerical test: Liquid Water}
\begin{figure}
  \centering
  \includegraphics[width=0.245\textwidth]{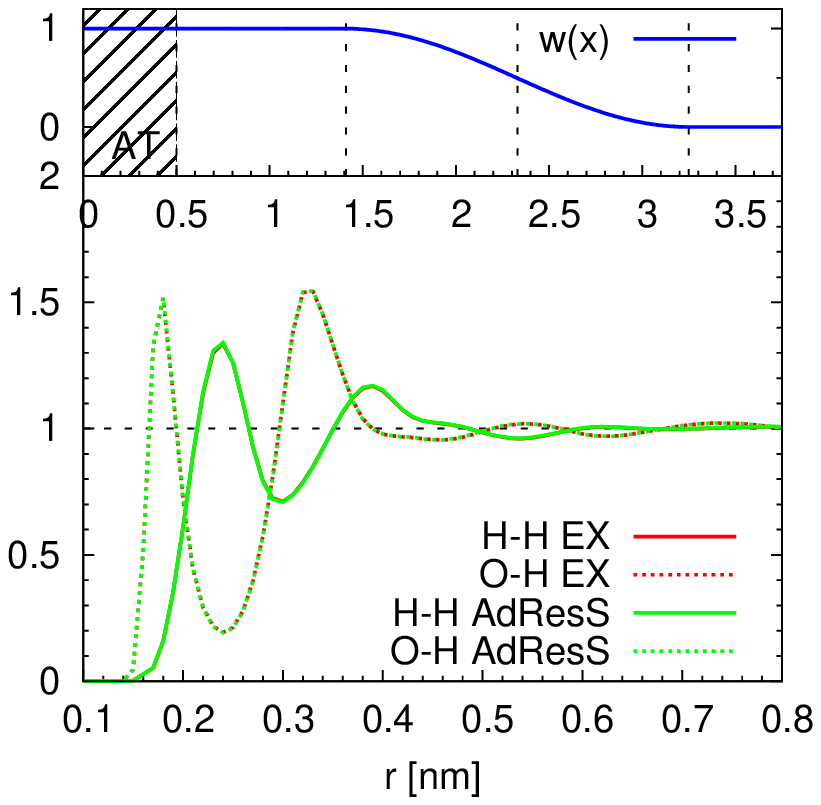}
  \includegraphics[width=0.245\textwidth]{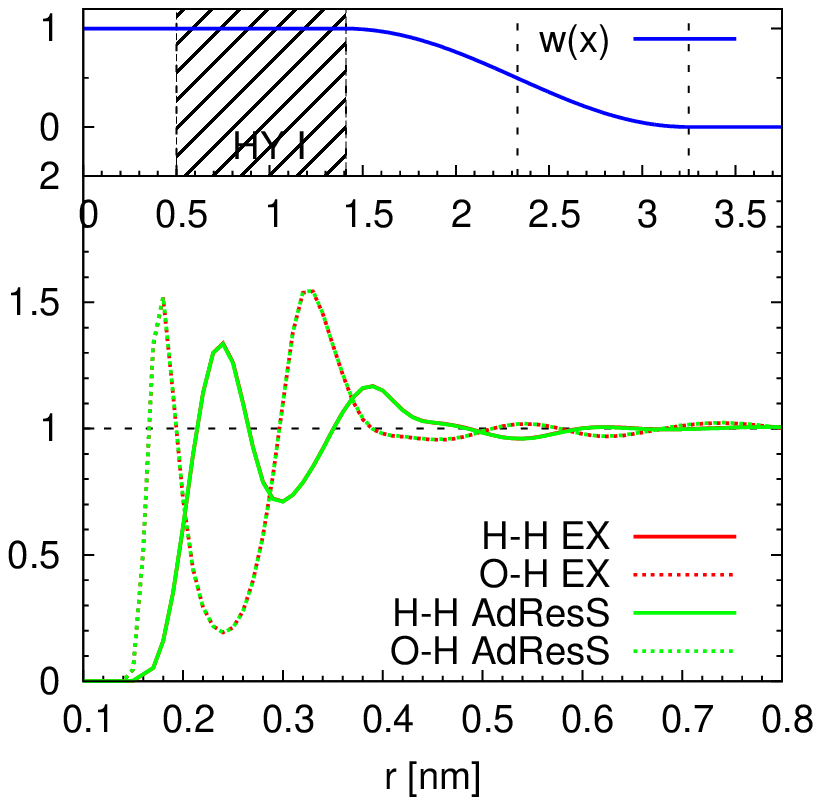}
  \includegraphics[width=0.245\textwidth]{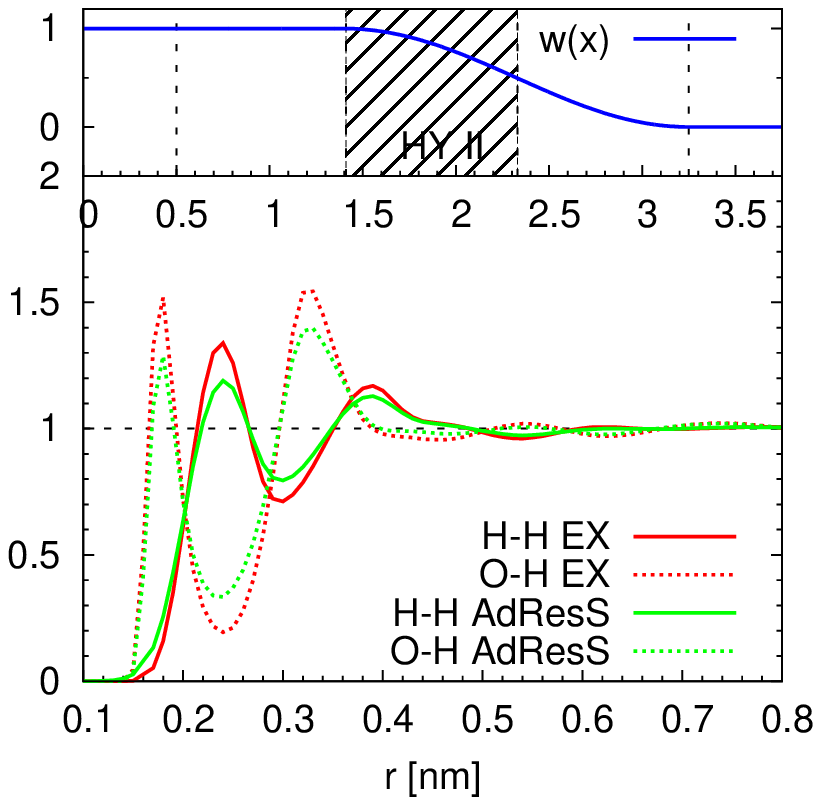}
  \includegraphics[width=0.245\textwidth]{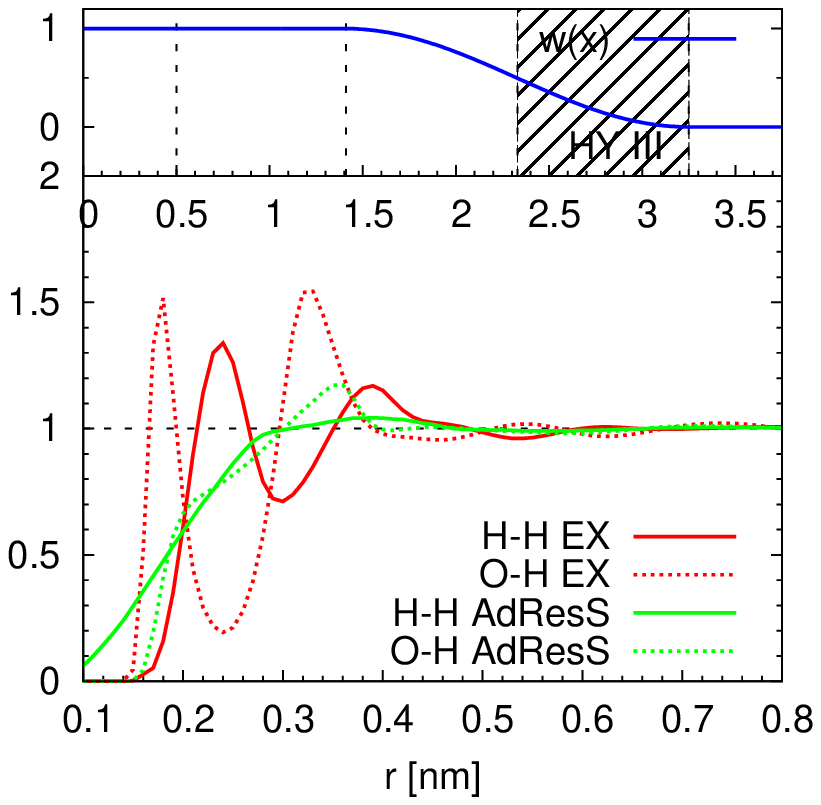}
  \caption{Local H-H and O-H $g(r)$'s.
    The red line is the curve corresponding to the reference explicit (all atom)
    simulation (EX).
    The curve obtained by employing the AdResS 
    method is represented in green.
    The hybrid region is equally
    divided into three parts: HY I, HY II and HY III, the widths of
    which are roughly equal to the cut-off radius, i.e. 9 \textsf{nm}.    
    The top part of each panel shows the region where the $g(r)$ is calculated,
    and the value of weighting function $w(x)$ there.
    From left to right, the panels correspond to the AT region, 
    the HY I subregion of $\Delta$,
    the HY II subregion of $\Delta$,
    and the HY III subregion of $\Delta$, respectively. It can be seen that beyond 0.5 \textsf{nm} the functions go to one, that is particle beyond this distance are uncorrelated; this is fully consistent with our hypothesis of Eq.\ref{eqn:p-1-1}.}
  \label{fig:tmp2a}
\end{figure}

\begin{figure}
  \centering
  \includegraphics[width=0.9\textwidth]{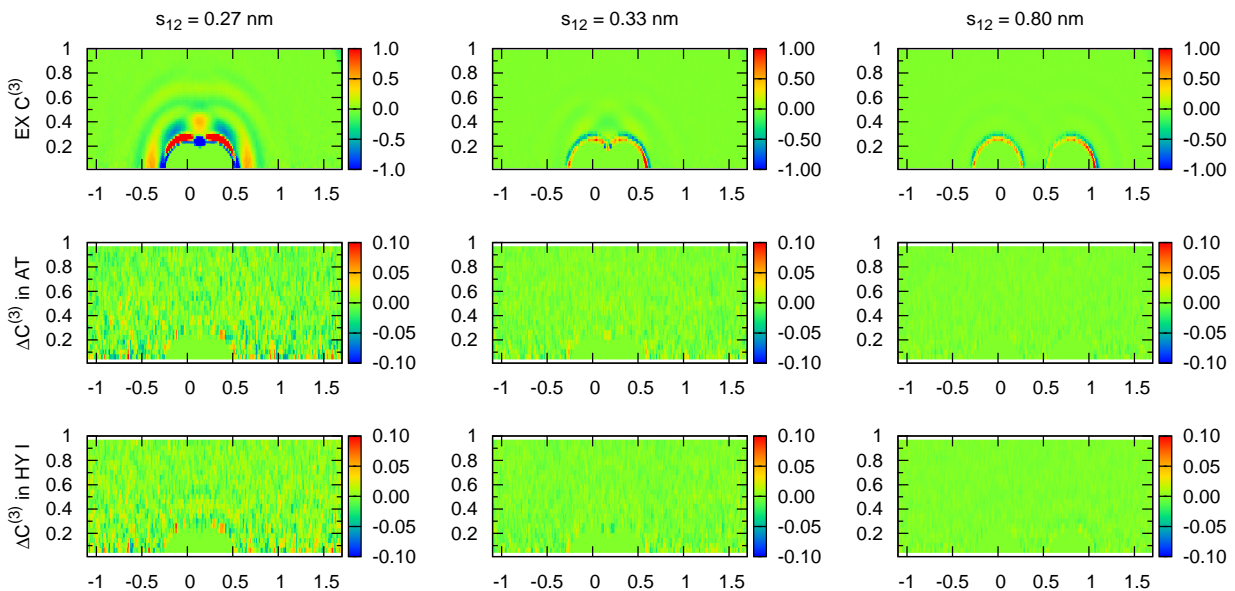}
  \caption{The 3-body correlation function $\corr$.  The 1st row corresponds to the reference all atom simulation (EX).
    The 2nd and 3rd row show the difference between the function calculated in AdResS in the AT, HY I
    regions, respectively, and the reference function of the full atomistic case.
    Different columns give different distances between
    the first two molecules: from left to right $s_{12} =
    0.27\,\textsf{nm},\ 0.33\,\textsf{nm},\  
    0.80\,\textsf{nm}$.  The horizontal axis corresponds to the variable $h_1$, while vertical
    axis is the variable $h_2$ (see the Appendix~\ref{app:3} for the
    definition of these variables).  The magnitude of the correlation
    and the differences are indicated by
    different colors.
  }
  \label{fig:tmp2b}
\end{figure}
The arguments we have used so far are not sufficient to infer that the probability $p(\vect x_1 , N_1)$,
as presented in the theoretical analysis, corresponds to that of a Grand Canonical distribution  (Eq.~\eqref{eqn:p}).
One must be very careful about any statement on $p(\vect x_1, N_1)$ 
due to the following three reasons:
\begin{enumerate}
\item There is no hope whatsoever that the assumed Boltzmann form of the distribution $p(\vect x_1|N_1; \vect x_2, N_2)$ can be numerically verified  for any realistic test system. 
\item Conditions \eqref{eqn:p-1-cond1} and \eqref{eqn:p-1-cond2}
are only necessary, and are in general not sufficient to guarantee
the correct distribution $p(\vect x_2, N_2 | N_1)$ in the hybrid region $\HY$.
\item Formally, the particle number density $p(N_1)$ is only a second-order approximation of the correct density $p_{\AT}(N_{1})$.
\end{enumerate}
However, the number distribution $p(N_1)$ has already
been proved to be numerically correct in Ref.~\onlinecite{prlgc, rdfcorr} for the relevant case of liquid water. In this section, we want to test, with a numerical simulation, to what extent, other quantities, e.g. the configurational distribution $p(\vect x_1) = \sum_{N_1}p(\vect x_1, N_1)$, are correct. 
Obviously it is a prohibitive task for any complex molecular system to determine the high-dimensional configurational distribution,
for this reason, instead, we study its marginal distributions up to the third order.
Here we report the H-O and H-H RDFs (see Fig.~\ref{fig:tmp2a}), and the 3-body COM correlation function
$\corr(\vect s_1, \vect s_2, \vect s_2)$ (see Fig.~\ref{fig:tmp2b}). For simulation
protocols and the definition of $\corr$, see the Appendix~\ref{app:2} and \ref{app:3}.
The first order marginal distribution, that is the molecular density profile and the second order marginal distribution, that is the COM RDF are not presented here, we address the readers to Ref.~\onlinecite{rdfcorr}. 
Here we focus on the more delicate RDFs which involve the atomistic accuracy, and on the third order COM three-body correlation function ($\corr$). From Fig.~\ref{fig:tmp2a} one can see that
the AdResS H-O and H-H RDFs are identical to those of 
the full atomistic reference system in the AT region.
Interestingly, despite the corrective force of the RDF is applied only to the COM RDF, also in the region HY I (that is close to the atomistic region) the H-O and H-H RDFs are the same as in the full atomistic case.
This implies that the AT region
is embedded in an environment where not only the COM-COM structural properties but also finer structural
properties are the same as for the atomistic resolution.
In the HY II and III regions, the AdResS results obviously deviate from those
of the full atomistic reference system, because the atomistic nature 
is decreasing as a molecule travels from HY I to HY III, so it must be expected 
that the orientational order is also fading away. In HY III, the H-O and H-H
RDFs are almost structureless.
Furthermore, we test $\corr$ in Fig.~\ref{fig:tmp2b} (for the definition of the three-body correlation function $\corr$ see Appendix~\ref{app:3}).
Interestingly, the three-body correlation is correctly reproduced
by the AdResS in the AT and HY I  regions. This is a further strong argument in favor of the fact that in AdResS the AT region is embedded in an environment that, at least up to the three-body COM correlation is the same as the full atomistic reference system.
In the HY II and HY III regions, the three-body correlation deviates from the
full atomistic reference, in fact since the various $g(r)$'s already deviates, one cannot expect a higher order correlation function to match (if not by chance) the full atomistic function of reference.
Another important information of Fig.~\ref{fig:tmp2a} and \ref{fig:tmp2b},
is that the hypothesis of short-range influence of a region over another employed in Eq.\ref{eqn:p-1-1} is numerically justified and thus we can state that for the case of liquid water at ambient conditions the AT region is only short-range
correlated with the rest of the system.

\section{WCA potential instead of atomistic-based coarse-grained potential: Coupling the atomistic region to a generic source of energy and particles}
The numerical test presented in the previous section shows how an atomistic model and its corresponding coarse-grained model, which resemble basic structural properties of the original full atomistic system, can be interfaced in an adaptive way and exchange molecules in a Grand Canonical fashion. However one may be tempted to think that the request of having a coarse-grained model which resembles as much as possible the structural and thermodynamic properties of a full atomistic model would be a necessary condition to have a proper exchange of particles between the different regions. If it was so, the idea of the adaptive scheme as a Grand Canonical tool would not be very general; our theoretical results instead suggest that it must be sufficient to impose some given conditions in the transition region to have a proper exchange of energy and particles between the small atomistic region and the large coarse-grained reservoir. 
This means that if the theoretical considerations are correct, one should be able to show numerically that the proper exchange of energy and molecules is independent from the molecular model used in the coarse-grained region; that is, the filter of the transition region makes the atomistic region a Grand Canonical ensemble. For this reason we have carried out a further numerical test where the molecules in the coarse-grained region interact with a generic WCA potential which assures that only the density, but not the radial distribution function, of a liquid governed by this potential is the same as that of the liquid water (for the simulation set up see the Appendix~\ref{app:2}).
\begin{figure}
  \centering
  \includegraphics[width=0.3\textwidth]{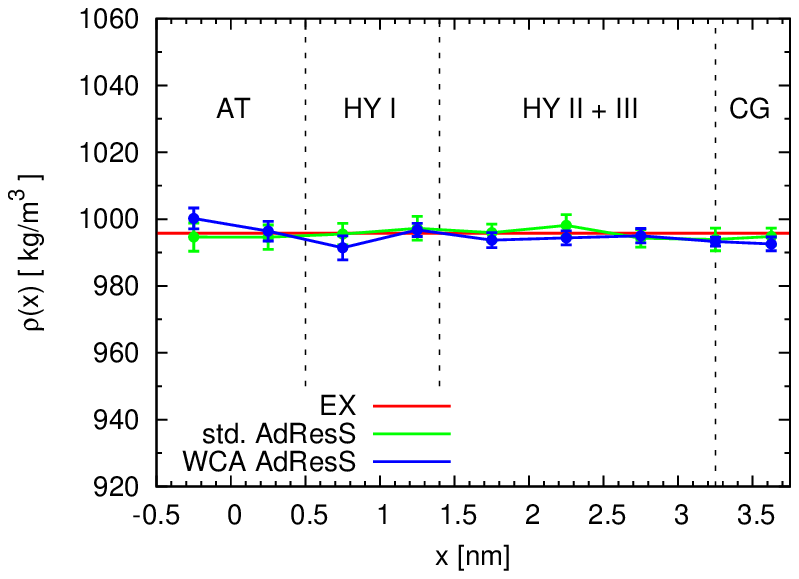}
  \includegraphics[width=0.3\textwidth]{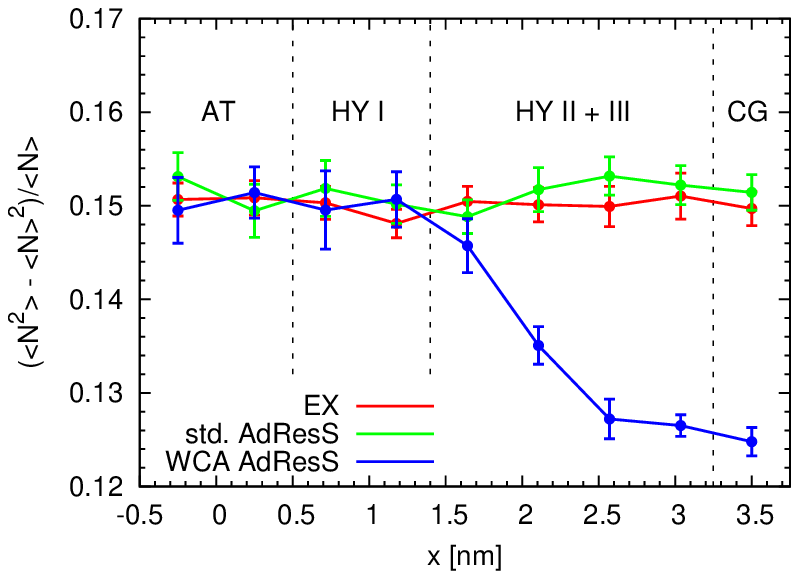}
  \caption{Density and fluctuation profile of the case of the WCA potential in AdResS.  The
    results are compared with the full atomistic reference (denoted by
    ``EX'') and the  standard AdResS (denoted by ``std. AdResS'').
    The error bars indicate the statistical error up to a confidential
    level of 95\%.
  }
  \label{fig:wca-den}
\end{figure}

\begin{figure}
  \centering
  \includegraphics[width=0.245\textwidth]{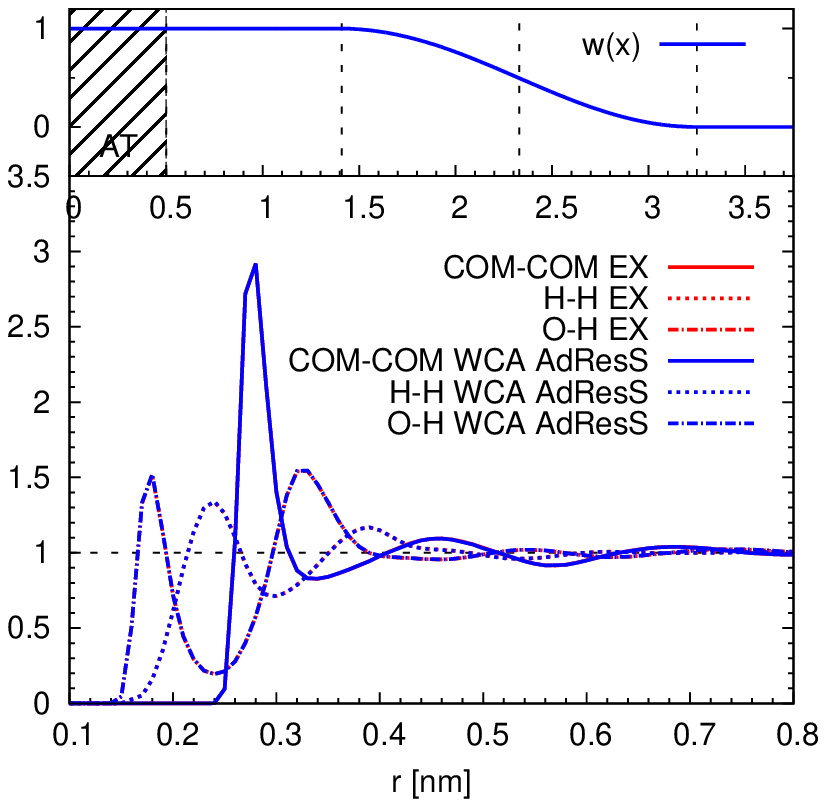}
  \includegraphics[width=0.245\textwidth]{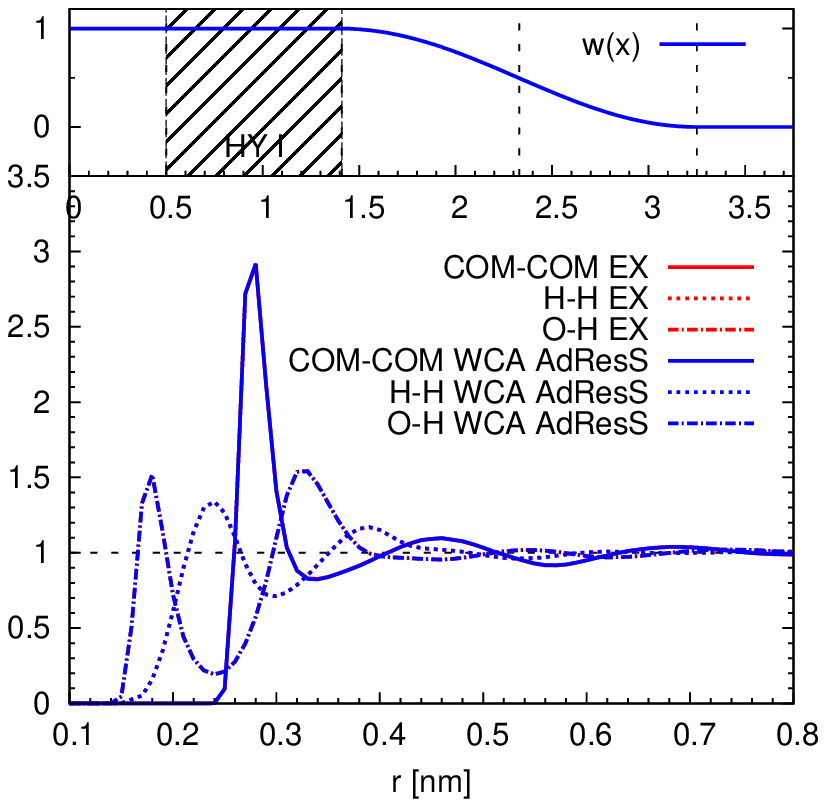}
  \includegraphics[width=0.245\textwidth]{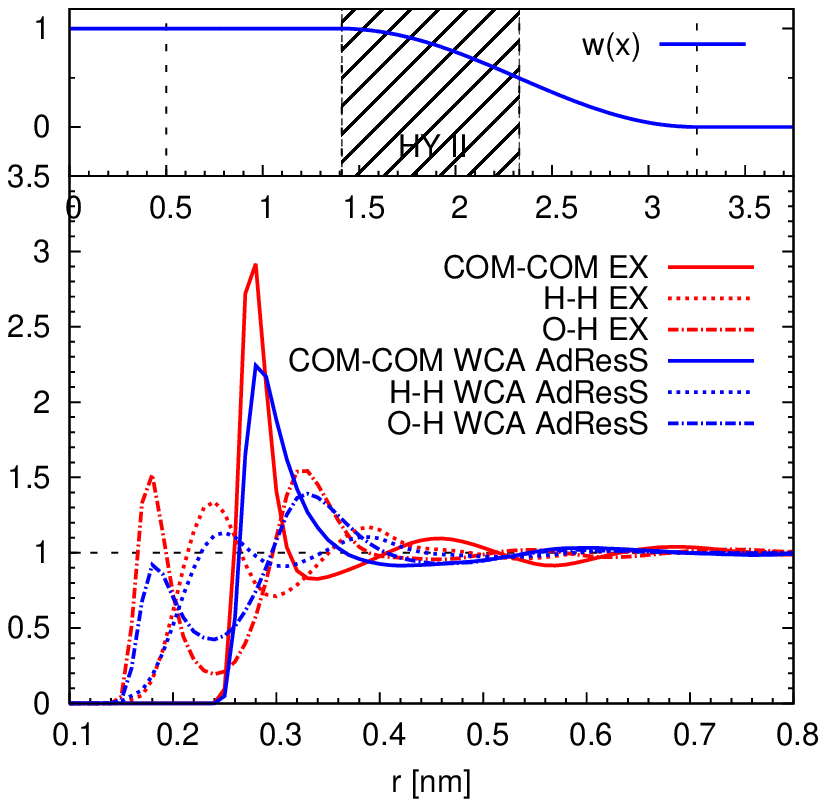}
  \includegraphics[width=0.245\textwidth]{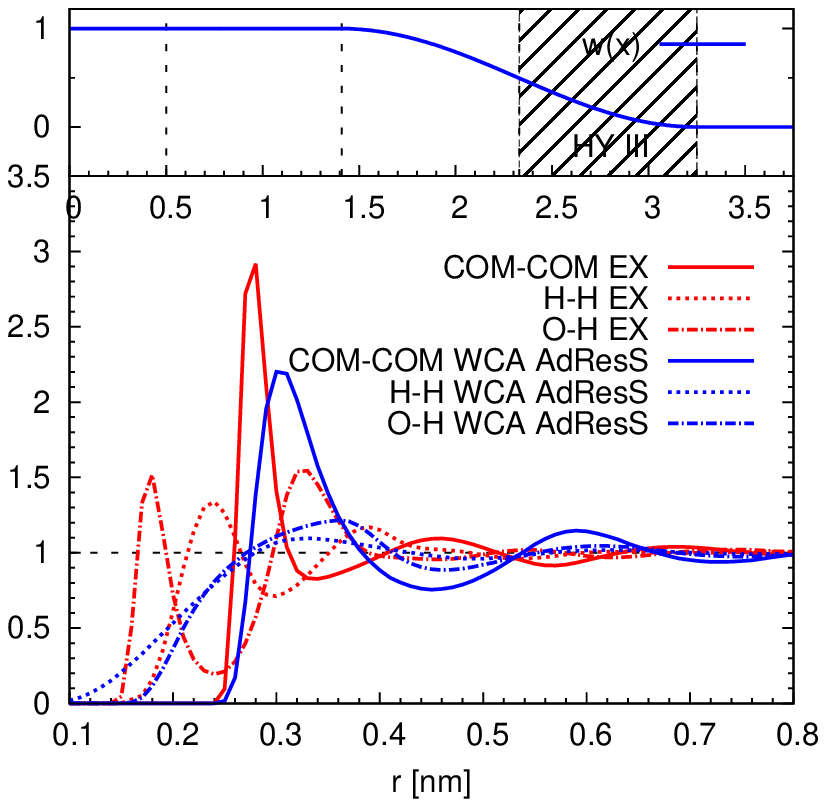}
  \caption{
    The COM-COM, H-O and H-H RDFs. 
    The red line is the curve corresponding to the reference explicit (all atom)
    simulation (EX).
    The curve obtained by employing the WCA AdResS 
    method is represented in blue.
    The hybrid region is equally
    divided into three parts: HY I, HY II and HY III, the widths of
    which are roughly equal to the cut-off radius, i.e. 9 \textsf{nm}.    
    The top part of each panel shows the region where the $g(r)$ is calculated,
    and the value of weighting function $w(x)$ there.
    From left to right, the panels correspond to the AT region, 
    the HY I subregion of $\Delta$,
    the HY II subregion of $\Delta$,
    and the HY III subregion of $\Delta$, respectively. 
  }
  \label{fig:wca-dist}
\end{figure}

\begin{figure}
  \centering
  \includegraphics[width=0.9\textwidth]{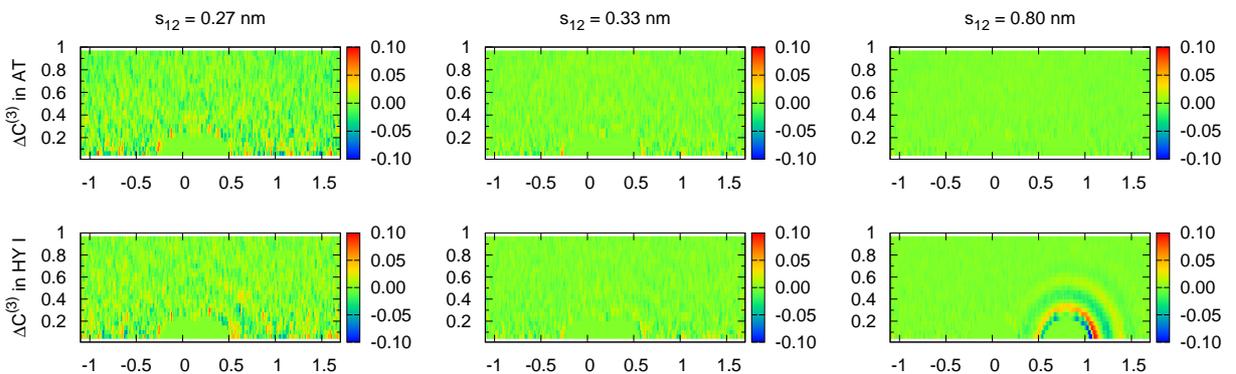}
  \caption{The 3-body correlation of WCA AdResS in the AT
    and HY I regions.
    The difference with respect to the full atomistic reference
    (see the fist line of Fig.~\ref{fig:tmp2b})
    is plotted in this figure. From left to right, the distance
    between the first two molecules are $s_{12} =
    0.27\,\textsf{nm},\ 0.33\,\textsf{nm},\  
    0.80\,\textsf{nm}$, respectively.
  }
  \label{fig:wca-g3}
\end{figure}

  The left plot of Fig.~\ref{fig:wca-den} shows that
  the density profile of the
  WCA AdResS is perfectly matching the all-atom reference system; satisfying agreement is found also in the AT and HY I regions of the right plot of Fig.~\ref{fig:wca-den} which shows the molecular
  number fluctuation over the whole system.
Fig.~\ref{fig:wca-dist} shows the COM-COM, H-O and H-H
    RDFs in the AT, HY I, HY II and HY III regions.
    All the RDFs are compared with full atomistic reference system (EX).
    Finally, Fig.~\ref{fig:wca-g3} compares the 3-body correlation $\corr$
    with the atomistic reference
    system (this latter reported in the first row of Fig.~\ref{fig:tmp2b}), and plots the difference. It must be noticed that in this case, on purpose, we want to investigate a {\it ``worst case scenario''} and thus we did not apply the RDF corrective force in the transition region. In this way we can understand whether or not the corrective action of the thermodynamic force only, is sufficient from the numerical point of view to reproduce the Grand Canonical-like environment for the AT region.
    According to our theoretical framework, in absence RDF corrective force only the first order of the distribution (i.e. the density profile)
    is corrected by the thermodynamic force in the $\HY$ region.
    However, numerically, Fig.~\ref{fig:wca-dist} shows a second order accuracy (correct RDFs) in the HY I region as well. Instead, the 3-body
    correlation in HY I deviates from the full atomistic reference; this is the price we pay for this ``worst case scenario'' set up. Nevertheless, the accuracy reached in the transition region is sufficient in order to have, at least up to the second order, the AT region of the WCA AdResS equivalent to the corresponding subregion in the full atomistic reference.
    Furthermore, numerical results  (Fig.~\ref{fig:wca-g3}) show that, in the AT region, even the 3-body correlation $\corr$ is nonetheless very accurate. 
  This means that, from the numerical point of view, even only \redc{the work of the thermodynamic force and that of the thermostat}, by allowing an exchange of particles under conditions of equilibrium of the chemical potential, is sufficient to assure the correct equilibrium of the atomistic region. The possible application of the corrective force on the RDF's in $\Delta$ in the case of liquid water is, in this case, numerically not required, but in case of necessity it would anyway represent a systematic tool for improving accuracy. 
  Summarizing, results reported in Figs.~\ref{fig:wca-den}, \ref{fig:wca-dist} and \ref{fig:wca-g3} show that in the atomistic region the accuracy at the third order and particle number fluctuations, for which the filter in the transition region is designed, are well preserved; that is the atomistic region exchange particle and energy with the reservoir in a proper Grand Canonical fashion (up to the level of approximation decided {\it a priori}). Being the molecules in the coarse-grained region a generic liquid model, we have proved numerically that indeed the adaptive technique employed here can be considered as a tool for general Grand Canonical Molecular Dynamics simulations. 

\section{Computational efficiency of AdResS as a Grand Canonical set up}
\begin{figure}
  \centering
  \includegraphics[]{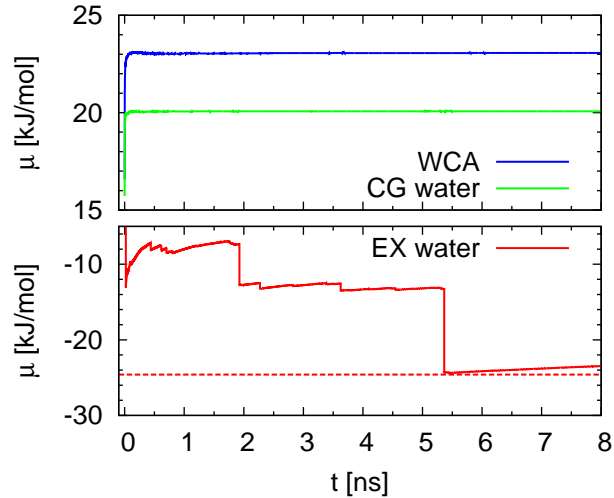}
  \caption{{The chemical potential calculated by the insertion particle method (IPM). The top panel shows the case of coarse-grained (CG) water and WCA liquid while the bottom panel shows the case of a full atomistic (EX) simulation. The horizontal dashed line in the bottom panel
        is used to guide the eyes and represent the ideal target. The plot shows
        the result that for EX water the IPM procedure does not converge within 8 \textsf{ns};
      in fact the convergence of the chemical potential means that the insertion of the molecule is properly done and subsequent part of the trajectory (with the associated IPM) can be used for ``statistical'' average in a Grand Canonical framework.}}
  \label{fig:tmp4}
\end{figure}

In order to show the computational efficiency of the AdResS method as a Grand Canonical set up, we have carried out a numerical experiment to compare the performance of our method in inserting and removing molecules from the atomistic region and that of a highly popular approach used in Molecular Dynamics, the insertion particle method (IPM) \cite{ipm}. The IPM is very efficiently implemented into the GROMACS code \cite{gromacs} and thus it is a natural choice as a Grand Canonical-like set up for a very large portion of molecular dynamics simulators. The experiment consists of the following: we have considered a large full atomistic system and applied the highly optimized IPM to insert molecules. From this study we extract the efficiency of the IPM in terms of computational resources required.
Next, we consider an adaptive system whose number of molecules, in the atomistic plus hybrid region, is at least the same as the full atomistic system considered (actually the for results presented here the size considered is even larger than that of the EX simulation); to this we add a rather large reservoir of coarse-grained particles (tests were carried out also with smaller reservoirs but the results were essentially the same as for the large reservoir). Also in this case the efficiency in measured terms of computational resources required.
The results show that the IPM procedure itself requires a rather high computational cost, to which it must be added long trajectories so that the insertion of { one molecule per time} converges and data can be then collected for real statistics. Instead, the adaptive approach performs, in less time and at a much lower computational cost, on-the-fly dynamical multiple insertions keeping the instantaneous equilibrium intact, which implies the additional advantage that data can be collected on-the-fly for real statistics at any time.
 Specifically, we used an equilibrium trajectory of length
8~\textsf{ns} where the coordinates of water molecules were recorded every
0.2~\textsf{ps}.  Using the IPM approach, in each frame, $10^5$ test particles were inserted
at random position of the system. For each insertion, $10^5$ randomly
chosen orientations were tried. The convergence is very
slow, even at 8~\textsf{ns} the insertion process is not satisfactorily converged (see Fig~\ref{fig:tmp4}), this means that up to this point data cannot be collected for analysis of physical properties.
Compared to IPM, AdResS leads a very
efficient insertion of particles: in each 1~\textsf{ps} time interval,
(on average) 23 water molecules entered into the atomistic region. Some of
them came immediately back, because of a unfavorable entering
configuration.  To investigate the proper insertion, we consider the
time interval of 1~\textsf{ns}, and find 832 out of 13824 molecules
were firstly in either the HY or CG region, and then travelled to the
AT region and stayed there for longer than 60~\textsf{ps}, which
indicates a mean square root displacement of roughly 1~\textsf{nm}
in the AT region. Moreover, while for the EX simulation one must avoid size effects by choosing reasonable large systems, in AdResS the AT region can actually be chosen to be much smaller than the one used here, thus enhancing the amount of efficiency. Finally, even if one can afford a rather large atomistic system, thus a faster convergence, the cost of the insertion procedure will increase enormously because it would require a larger sampling due to the increased number of possibility to allocate a particle in the liquid. 
These results in terms of computational resources are summarized in Table~\ref{tab:tmp1}.
\begin{table}
  \centering
  \begin{tabular}{l|l|l|l|l}
    & Sys. size ($\textsf{nm}^3$)
    & Traj. length (\textsf{ns})
    & CPU time (hours)\\    \hline
    AdResS   &$29.9\times3.7\times3.7$ & 1 & 3.1\\
    EX water & $3.7\times3.7\times3.7$ & 8 & 4.5 (traj.) + 36.7 (IPM)\\
  \end{tabular}
  \caption{Comparison of computational efficiency.
      The AdResS simulation contains 13824 molecules, while the full atomistic (EX)
      water simulations contain 1728 molecules. In the AdResS simulation,
      the size of the AT + HY regions ($6.7\times3.7\times3.7\textsf{nm}^3$) is even larger than that of the EX water
      simulation box, so that we are actually in a {\it ``worst scenario''} situation. Moreover simulations with smaller reservoirs  gave essentially the same results reported for this system.
      ``Sys. size'' means the size of the box used in the simulation.
      ``Traj. length'' means the equilibrium trajectory
      used for the particle insertion in the IPM approach. Along the trajectories, frames
      of configurations were recorded every 0.2~\textsf{ps}.
      ``CPU time'' the is wall clock time spend on the simulation.
      For the particle insertion simulation, the CPU time is counted by two
      parts: the time of generating the equilibrium trajectories (traj.) (not expensive)
      and the time required by the IPM procedure (very expensive). Moreover, not only the insertion procedure is expensive on itself, but even after $8\textsf{ns}$ the insertion process has not actually converged. Simulations on longer time scales show that the full convergence is actually never reached, which suggests that the insertion procedure for a molecule as simple as water is not fully rigorous from the physical point of view.
    }
  \label{tab:tmp1}
\end{table}
{Moreover, in previous work, Praprotnik {\it et al.} \cite{delgado2009coupling} have shown that the coarse-grained region of AdResS can be easily coupled to the continuum; this has the non trivial advantage that the insertion of a sphere in a liquid of spheres by IPM is by far much simpler than inserting an atomistically structured molecule in a liquid of atomistic molecules. Fig.\ref{fig:tmp4} shows the extremely fast convergence for inserting a particle in our ``reservoir'' of spherical particles compared to that of a full atomistic system. This means that in the long term perspective the part of the algorithm of Praprotnik {\it et al.} could be merged with our current approach and thus allow for even smaller sizes of the reservoir region. However, this extension, though conceptually straightforward goes beyond the scope of this paper. In general, there have been a large number of developments for the GC idea based on the { ``grow-in''} and { ``shrink-out''} MD approach (see e.g. Lynch and Petitt \cite{pet5} and Eslami and M\"uller-Plathe~\cite{flo}) however, all of them  share the same basic principle and face the same problem of the IPM. This numerical experiment shows that despite the need of a reservoir, our computation is by far more efficient than any other technique so far used.} \redc{The comparison between the performance of AdResS and IPM as a Grand Canonical set up, leads to the natural question of whether or not our method can be employed to calculate 
the excess chemical potential which is the primary purpose for which IPM is used. If this is possible, then AdResS may represent a more efficient computational tool than IPM also
for calculating such a quantity. In Appendix~\ref{app:4} we provide a practical scheme for using AdResS as a tool to calculate the chemical potential and show that the results are in rather satisfying agreement with those obtained by IPM.}

\section{Conclusions}
We have shown a detailed theoretical analysis about the validity, the limitations and meaning of the AdResS method within the framework of a Grand Canonical ensemble.
Using strict formal arguments, we have demonstrated the \redc{role of the thermodynamic force and of the thermostat in balancing the difference in chemical potential due to the different resolutions in space.}
  The theoretical results, derived under the assumption of thermodynamic limit, are then checked with several numerical tests. These latter represent a {\it "worst scenario''} case since the conditions of simulation are not ideal as in the analytical procedure.
{Next}, we have shown that, under given hypothesis, $p(N_{1})$, the probability distribution in the atomistic region of AdResS, is equivalent to that in the same region in a full atomistic simulation up to the accuracy of second order. We have further strengthen our hypothesis and conclusions by carrying out a numerical experiment for the case of liquid water at ambient conditions where we could even go beyond the COM RDF. 
A further numerical experiment was made where the atomistic-based coarse-grained model was substituted by a generic liquid of spheres interacting via WCA potential. Results show that the accuracy in the atomistic region is the same as that of AdResS based on a coarse-grained model derived from the full atomistic reference system; this strengthen the idea of AdResS as a tool for coupling the atomistic region to a generic source of energy and particles in a Grand Canonical manner. {Finally we validate the efficiency of our method as a Grand Canonical set up comparing its computational performance to that of well established and largely used alternative methods.} 
In conclusion, these results provide a solid theoretical basis to explain the numerical reliability of the AdResS as an effective Grand Canonical set up and provide basis for further formal and numerical development of the adaptive idea in terms of matching probability distributions.

\section*{Acknowledgement}
This work was partially supported by the Deutsche Forschungsgemeinschaft (DFG) with the Heisenberg grant provided to L.D.S (grant code DE 1140/5-1) and by ITN Nanopoly provided to H.W and C.S.
It was also supported in part by the National Science Foundation under Grant No. NSF PHY11-25915.

\section*{References}
\bibliography{ref}{}
\bibliographystyle{unsrt}

\appendix

\section{The number probability $p(N_1)$}\label{app:1}
In this section, we define at which level of approximation $p(N_{1})$, in AdResS is the same as in a full atomistic simulation. 
Let us consider the marginal distribution $p(N_1)$
\begin{align}\nonumber
  p(N_1)
  &=
  \frac{N!}{N_1!N_3!}
  \int
  \dd d\vect x_1\dd d\vect x_3  \:
  \frac{1}{Q(N,V,T) N!}
  e^{-\beta
    [\mathcal H_{N_1}^{\AT}(\vect x_1) +
    \mathcal H_{N_3}^{\CG}(\vect x_3) +
    N_1\omega_0]}\\\nonumber
  &=
  \frac{N!}{N_1!N_3!}
  \frac{N_1!N_3!}{Q(N,V,T) N!}
  e^{-\beta N_1\omega_0}
  \bigg[
  \frac1{N_1!}
  \int
  \dd d\vect x_1
  e^{-\beta \mathcal H_{N_1}^{\AT}(\vect x_1)}
  \bigg]
  \bigg[
  \frac1{N_3!}
  \int
  \dd d\vect x_3
  e^{-\beta \mathcal H_{N_3}^{\CG}(\vect x_3)}
  \bigg]  \\\label{eqn:pn1}
  &=
  \frac
  {
    e^{-\beta N_1\omega_0}
    Q_{\AT}(N_1,V_1,T) Q_{\CG}(N-N_1,V_3,T)
  }
  {
    \sum_{n_1}
    e^{-\beta n_1\omega_0}
    Q_{\AT}(n_1,V_1,T) Q_{\CG}(N-n_1,V_3,T)
  }
\end{align}
where $n_1$ is the summation variable that goes from 0 to $N$.
In any case, when $n_1$ is not much smaller than $N$ (i.e. out of the range of the
thermodynamic limit), the statistic is not relevant,
and can be safely neglected.
The marginal number distribution of the full atomistic simulation,
which the AdResS simulation should
reproduce is
\begin{align}
  p_{\AT}(N_1)
  &=
  \frac
  {
    Q_{\AT}(N_1,V_1,T) Q_{\AT}(N-N_1,V_3,T)
  }
  {
    \sum_{n_1}
    Q_{\AT}(n_1,V_1,T) Q_{\AT}(N-n_1,V_3,T)
  }  
\end{align}
We want to calculate the difference between $p(N_1)$ and $p_{\AT}(N_1)$; that is the difference between the particle probability in the atomistic region of AdResS ($p(N_{1})$), and the particle probability in a full atomistic simulation in the region equivalent to the AT region of AdResS.
In order to do so we proceed with the following technical operation: multiply the numerator of $p(N_1)$ by the denominator of
$p_{\AT}(N_1)$ (denoted by $T_1$),
and multiply the numerator of $p_{\AT}(N_1)$
by the denominator of $p(N_1)$ (denoted by $T_2$). It follows:
\begin{align}
  T_1
  &=
  e^{-\beta N_1\omega_0}
  Q_{\AT}(N_1,V_1,T) Q_{\CG}(N-N_1,V_3,T)
  \times
  \sum_{n_1}
  Q_{\AT}(n_1,V_1,T) Q_{\AT}(N-n_1,V_3,T)\\
  T_2
  &=
  Q_{\AT}(N_1,V_1,T) Q_{\AT}(N-N_1,V_3,T)
  \times
  \sum_{n_1}
  e^{-\beta n_1\omega_0}
  Q_{\AT}(n_1,V_1,T) Q_{\CG}(N-n_1,V_3,T)
\end{align}
The difference between $T_1$ and $T_2$ is basically the difference
between $p(N_1)$ and $p_{\AT}(N_1)$.
Calculating $T_1$:
\begin{align}\nonumber
  T_1
  &=
  \sum_{n_1}
  e^{-\beta N_1\omega_0}
  Q_{\AT}(n_1,V_1,T)\,
  Q_{\AT}(N_1,V_1,T)\,
  Q_{\AT}(N-n_1,V_3,T)\,
  Q_{\CG}(N-N_1,V_3,T) \\\label{eqn:t1-1}
  &=
  \sum_{n_1}
  \exp
  \big\{-\beta
  \big[
  \omega_0N_1 +
  A_{\AT}(n_1,V_1,T) +
  A_{\AT}(N_1,V_1,T) +
  A_{\AT}(N-n_1,V_3,T) +
  A_{\CG}(N-N_1,V_3,T)
  \big]
  \big\}
\end{align}
Notice that the free energy is extensive.
By applying Euler's theorem~\cite{tuckeman2010statistical} we thus find 
\begin{align}\label{eqn:Aat-0}
  A_{\AT}(N-n_1,V_3,T)
  &= V_3 A_{\AT}(\frac{N-n_1}{V_3},1,T)
  = V_3 A_{\AT}(\rho_0 + \frac{\hat N_1 - n_1}{V_3},1,T)\\\label{eqn:Acg-0}
  A_{\CG}(N-N_1,V_3,T)
  &= V_3 A_{\CG}(\frac{N-N_1}{V_3},1,T)
  = V_3 A_{\CG}(\rho_0 + \frac{\hat N_1 - N_1}{V_3},1,T),
\end{align}
where $\hat N_1$ is the number of molecules in the atomistic region
at the wished density $\rho_0$ of equilibrium, namely
$\hat N_1 = \rho_0V_1 = N - \rho_0 V_3$. As we work under the hypothesis that the atomistic region is much smaller than the whole system, it
is reasonable to assume $N_1\ll N$. Moreover we have found that, in the thermodynamic limit, $N_{1}\sim\hat N_1\sim\bar N_1$ which carries over to the running index $n_{1}$  in the grand canonical partition function that is non-negligible only if $n_{1}\sim\hat N_1\sim\bar N_1$. Hence 
$A_{\AT}(\rho_0 + (\hat N_1 - n_1)/{V_3},1,T)$ and
$A_{\CG}(\rho_0 + (\hat N_1 - N_1)/{V_3},1,T)$ can be regarded as
the free energies per \emph{unit} volume, with
the density perturbed from $\rho_0$ by  $(\hat N_1 - n_1)/{V_3}$
and $(\hat N_1 - N_1)/{V_3}$, respectively.
These perturbations originate from the fluctuating
number of molecules in the atomistic region: when the number
deviates from $\hat N_1$, molecules enter into the coarse-grained
region and the density is perturbed.
In the thermodynamic
limit, $\hat N_1$ is comparable to $n_1$ and $N_1$, and all of them are
supposed to be much much smaller than $N$ while $V_3$ is of the same order of magnitude of $V$ (the large CG region compared to the small AT region). As a consequence the perturbations
$(\hat N_1 - n_1)/{V_3}$ and $(\hat N_1 - N_1)/{V_3}$ are small, compared to $\rho_0 = N/V$ and
Taylor expanding the free energies \eqref{eqn:Aat-0} and \eqref{eqn:Acg-0}
about $\rho_0$ yields
\begin{align}
  A_{\AT}(\rho_0 + \frac{\hat N_1 - n_1}{V_3},1,T)
  &= A_{\AT}(\rho_0,1,T)
  +\mu_{\AT}
  \Big(
  \frac{\hat N_1 - n_1}{V_3}
  \Big)
  +
  \frac1{2\rho_0^2\,\kappa_{\AT}}
  \Big(
  \frac{\hat N_1 - n_1}{V_3}
  \Big)^2
  + o(1) \\
  A_{\CG}(\rho_0 + \frac{\hat N_1 - N_1}{V_3},1,T)
  &= A_{\CG}(\rho_0,1,T)
  +\mu_{\CG}
  \Big(
  \frac{\hat N_1 - N_1}{V_3}
  \Big)
  +
  \frac1{2\rho_0^2\,\kappa_{\CG}}
  \Big(
  \frac{\hat N_1 - N_1}{V_3}
  \Big)^2
  + o(1)
\end{align}
(We use the Landau notation $o(1)$ to collect all terms that are asymptotically negligible, bearing in mind that in the thermodynamic limit $N_{1}\sim\hat{N}_{1}$ ;  see also Ref.~\onlinecite{rdfcorr}).
By using the conditions \eqref{eqn:mu-eq} and \eqref{eqn:kappa-eq},
we have $T_1$
\begin{align}\nonumber
  T_1
  = \,&
  \sum_{n_1}
  \exp
  \Big\{-\beta
  \Big[
  A_{\AT}(n_1,V_1,T) +
  A_{\AT}(N_1,V_1,T) +
  A_{\AT}(\rho_0V_3,V_3,T) +
  A_{\CG}(\rho_0V_3,V_3,T) \\
  \,&+(\mu_\AT + \mu_\CG)\hat N_1
  -\mu_{\AT}\,(N_1 + n_1) +
  \frac{V_3}{2\rho_0\, \kappa_{\AT}}
  \Big[
  \Big(
  \frac{\hat N_1 - n_1}{V_3}
  \Big)^2
  +
  \Big(
  \frac{\hat N_1 - N_1}{V_3}
  \Big)^2
  \Big]
  + o(1)
  \Big]
  \Big\}
\end{align}
Similarly, we calculate $T_2$:
\begin{align}\nonumber
  T_2
  &=
  \sum_{n_1}
  e^{-\beta n_1\omega_0}
  Q_{\AT}(n_1,V_1,T)\,
  Q_{\AT}(N_1,V_1,T)\,
  Q_{\CG}(N-n_1,V_3,T)\,
  Q_{\AT}(N-N_1,V_3,T) \\\nonumber
  &=
  \sum_{n_1}
  \exp
  \big\{-\beta
  \big[
  \omega_0n_1 +
  A_{\AT}(n_1,V_1,T) +
  A_{\AT}(N_1,V_1,T) +
  A_{\CG}(N-n_1,V_3,T) +
  A_{\AT}(N-N_1,V_3,T)
  \big]
  \big\}
\end{align}
with similar expansions and by the conditions Eq.~\eqref{eqn:mu-eq}
and \eqref{eqn:kappa-eq}, we have:
\begin{align}\nonumber
  T_2
  = \,&
  \sum_{n_1}
  \exp
  \Big\{-\beta
  \Big[
  A_{\AT}(n_1,V_1,T) +
  A_{\AT}(N_1,V_1,T) +
  A_{\AT}(\rho_0V_3,V_3,T) +
  A_{\CG}(\rho_0V_3,V_3,T) \\
  \,&+(\mu_\AT + \mu_\CG)\hat N_1
  -\mu_{\AT}\,(N_1 + n_1) +
  \frac{V_3}{2\rho_0\, \kappa_{\AT}}
  \Big[
  \Big(
  \frac{\hat N_1 - n_1}{V_3}
  \Big)^2
  +
  \Big(
  \frac{\hat N_1 - N_1}{V_3}
  \Big)^2
  \Big]
  + o(1)
  \Big]
  \Big\}
\end{align}
Since the difference between the probabilities
$p(N_1)$ and $p_{\AT}(N_1)$ is basically the difference between $T_1$
and $T_2$ renormalized by the partition functions of the distributions,
by
invoking Laplace's method again,
we see $p(N_1)$ and $p_{\AT}(N_1)$ agree up to
the second leading
order with respect to the density perturbations
$(\hat N_1 - n_1)/{V_3}$
and $(\hat N_1 - N_1)/{V_3}$,
in the limit $V_{3}\to\infty$.


\section{Simulation protocols}
\label{app:2}
The testing system contained 3456 SPC/E \cite{berendsen1987missing}
water molecules in a $7.50\textsf{nm}\times 3.72\textsf{nm}\times
3.72\textsf{nm}$ periodic box. The system was divided along the $x$ direction
into one atomistic region of width $1.00\textsf{nm}$ and one
coarse-grained region of width $1.00\textsf{nm}$ connected by two
hybrid region of width $2.75\textsf{nm}$.
The hybrid region was equally divided into three sub-regions (with width $\sim 0.9\textsf{nm}$)
to calculate the local RDFs.
These sub-regions are called HY I, HY II and HY III, from the AT side to the CG resolution side.
The simulation was made at
room temperature of $300\textsf{K}$. The time step was $\Delta t =
0.002\textsf{ps}$. The cut-off radius $r_{c}$ used for all interactions was
$0.90\textsf{nm}$. The electrostatic interaction method used for the
atomistic region was the reaction field method. All simulations were
performed by MD simulation software Gromacs \cite{gromacs}
and VOTCA \cite{ruehle2009versatile}.
In the WCA-based simulation, the interaction potential between particles in the coarse-grained region is given by
\begin{align}
  U(r) = 4\varepsilon
  \big[
  \big(\frac\sigma r\big)^{12}
  -
  \big(\frac\sigma r\big)^{6}
  \big]
  + \varepsilon
  \qquad  r\leq 2^{1/6}\sigma
\end{align}
The parameters used in this study are $\varepsilon = 0.65$~\textsf{kJ/mol}
and $\sigma = 0.30$~\textsf{nm}. These parameters are chosen such that
the soft core of WCA is roughly the same as the structure-based coarse-grained
water model.
Other parameters used in the simulation are the
same as the standard AdResS.

\section{{The third order correlation function}}
\label{app:3}
{
\begin{figure}
  \centering
  \includegraphics[width=0.3\textwidth]{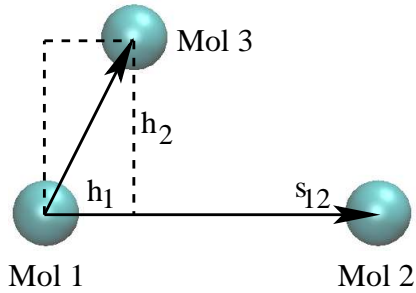}
  \caption{The schematic plot of the 3-body correlation function.}\label{fig:tmp3}
\end{figure}
We consider the following definition of three-body correlation function of a molecular system:
\begin{align}
  \corr (\vect s_1, \vect s_2, \vect s_3)
  =
  \frac1{\rho(\vect s_1)\rho(\vect s_2)\rho(\vect s_3)}
  \big\langle
  (\hat\rho(\vect s_1) - \rho(\vect s_1))\cdot
  (\hat\rho(\vect s_2) - \rho(\vect s_2))\cdot
  (\hat\rho(\vect s_3) - \rho(\vect s_3))
  \big\rangle
\end{align}
where the transient density distribution $\hat\rho(\vect s)$ and the
average density distribution $\rho(\vect s)$ are given
by:
\begin{align}
  \hat\rho(\vect s) = \sum_{i=1}^N\delta(\vect s - \vect r_i)
  \quad\textrm{and}\quad
  \rho(\vect s) = \big\langle \hat\rho(\vect s)\big\rangle.
\end{align}
By assuming that the system is homogeneous, one easily can show that
\begin{align}
  \corr (\vect s_1, \vect s_2, \vect s_3)
  =
  \frac1{\rho^3}\,\rho(\vect s_1, \vect s_2, \vect s_3) -
  \frac1{\rho^2}\,\big[
  \rho(\vect s_1, \vect s_2) +
  \rho(\vect s_1, \vect s_3) +
  \rho(\vect s_2, \vect s_3)
  \big] + 2
\end{align}
where
\begin{align}
  \rho(\vect s_1, \vect s_2)
  =
  \big\langle
  \hat\rho(\vect s_1)\,  \hat\rho(\vect s_2)
  \big\rangle
  \quad\textrm{and}\quad
  \rho(\vect s_1, \vect s_2, \vect s_3)
  =
  \big\langle
  \hat\rho(\vect s_1)\,  \hat\rho(\vect s_2)\, \hat\rho(\vect s_3)
  \big\rangle  
\end{align}
Since $g (\vect s_1, \vect s_2) = \rho(\vect s_1, \vect s_2)/\rho^2$
is the two-body radial distribution function, it follows:
\begin{align}
  \corr (\vect s_1, \vect s_2, \vect s_3)
  =
  \frac1{\rho^3}\,\rho(\vect s_1, \vect s_2, \vect s_3) -
  \big[
  g(s_{12}) +
  g(s_{13}) +
  g(s_{23})
  \big] + 2
\end{align}
where, for example, $s_{12} = \vert \vect s_1 -\vect s_2\vert$.

Since we assume the system to be homogeneous, the visualization of this high-dimensional
distribution is simplified as follows. We first fix the distance between molecule 1 and 2 (Mol 1 and 2), then plot the correlation function with
respect to the position of the third molecule (Mol 3, see
Fig.~\ref{fig:tmp3}).  The position of Mol 3 can be described by
the variables $h_1$ and $h_2$, which are
the projection of Mol~3's position on the to the
axis $\vect s_{12}$ defined by Mol 1 and 2.}
\section{Estimate of the excess chemical potential by AdResS}
\label{app:4}
 \redc{In this paper we have proven that the work of a thermodynamic force and the work of the thermostat in the transition region, which are the ingredients needed to provide density equilibrium across the system, correspond to the difference of the total (kinetic and excess) chemical potential between the atomistic and reservoir region. Thus in principle the calculation of the total work done by the thermodynamic force and that of the thermostat would allow for the determination of, at least, differences of total chemical potential, and, when the kinetic part is properly removed, of differences in the excess chemical potential.
While the calculation of the work of the thermodynamic force, $\omega_\thf$ implies simply the straightforward calculation of the integral of such a force over the transition region, the work of the thermostat , $\omega_\thermo$, for allowing the proper change of resolution of the molecules across the transition region is somewhat a much more delicate issue. In essence, the calculation of the chemical potential, would require the accurate determination of $\omega_\thermo^\ext$, which unfortunately cannot be done in a direct and efficient way. However, we have derived a procedure by which such a quantity can be calculate in an indirect way without additional computational efforts in a standard AdResS simulation, in the following part we illustrate such a procedure.}
\subsection{Auxiliary Hamiltonian Approach}
\begin{figure}
  \centering
  \includegraphics[]{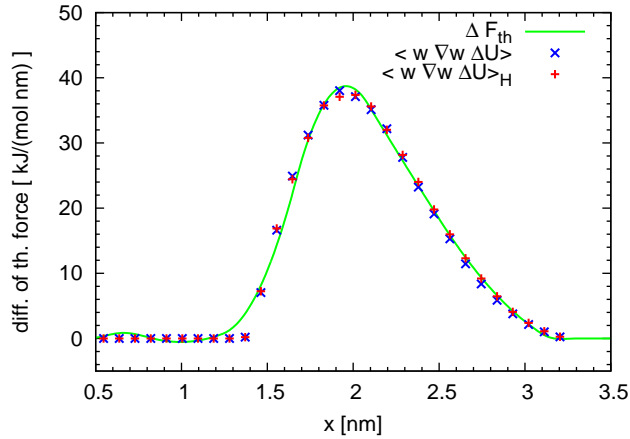}
  \caption{\redc{The difference between the thermodynamic force of the
    potential interpolation-based scheme and force interpolation of the standard AdResS.
    The green solid line shows the difference of the thermodynamic
    forces, i.e. $\Delta{\vect F}_\thf={\vect F}_\thf^\hadress-{\vect F}_\thf$.
    The blue and red symbols indicate the quantity $\langle w\nabla w(U^{\AT}-U^{\CG})\rangle$ calculated in the two different approaches (blue corresponds to the standard AdResS while red to the Hamiltonian-based approach). The equivalence between the two approaches is suggested by the fact that $\Delta{\vect F}_\thf$ corresponds to $\langle w\nabla w(U^{\AT}-U^{\CG})\rangle$, and that this latter is actually the same independently from the approach used to calculate it.}
  }
  \label{fig:diff-thf}
\end{figure}

\redc{
We define a system characterized by an atomistic region, a transition region and a coarse-grained (reservoir) region, as in a standard AdResS. We couple the different resolutions of the system via a Hamiltonian where the total potential is given by the interpolation of atomistic and coarse-grained potential in the same fashion of the force interpolation of standard AdResS:
\begin{equation}
U=\sum_{i<j}w_i w_j\,U^{\AT}_{ij}+\sum_{i<j}(1 - w_i w_j)\,U^{\CG}_{ij}.
\label{hinter}
\end{equation}
Next we determine the thermodynamic force for this system which leads to the same equilibrium density as that of the standard AdResS. We indicate such as force as ${\vect F}_\thf^\hadress$, and the superscript H is used to distinguish it from the thermodynamic force calculated  with the standard AdResS, which we have indicated as ${\vect F}_\thf$. We have now the following situation; in the approach based on the Hamiltonian, the changing of resolution in the transition region at equilibrium is provided by  the work of ${\vect F}_\thf^\hadress$ (here indicated as $\omega_\thf^\hadress$) and by the work of the thermostat to thermalize the reintroduced/removed  degrees of freedom $\omega_\dof$. Instead, in the standard AdResS, we have the same equilibrium density of the Hamiltonian approach, but the change of resolution is provided by ${\vect F}_\thf$, $\omega_\dof$ and $\omega_\thermo^\ext$. Since we have the same equilibrium  and $\omega_\dof$ is the same in both cases, it follows that:
\begin{equation}
\omega_\thermo^\ext=\omega_\thf^{H}-\omega_\thf.
\label{extraq}
\end{equation}
The limitation of this procedure would be the fact that $\omega_\thermo^\ext$ (and thus the chemical potential) could not be calculated within a standard AdResS simulation and one would require an auxiliary simulation, although the calculation is anyway not computationally demanding. However, based on an extended number of numerical tests, we could develop this idea further and
{calculate it by the standard AdResS simulation.}
To do so we have numerically studied the quantity $\Delta{\vect  F}_\thf={\vect F}_\thf^\hadress-{\vect F}_\thf$ for an extended number of systems, and proved that this is equal to $\langle w\nabla w(U^{\AT}-U^{\CG})\rangle$ (in Fig.\ref{fig:diff-thf}, is reported the case of liquid water). This quantity can be easily calculated without additional costs in a standard AdResS simulation, so $\omega_\thermo^\ext$ is calculated by
  \begin{align}
    \omega_\thermo^\ext
    = \int\dd d\vect r\,\Delta\vect F_\thf
    = \int\dd d\vect r\,\langle w\nabla w(U^{\AT}-U^{\CG})\rangle
  \end{align}
  The quantity $-\langle w\nabla w(U^{\AT}-U^{\CG})\rangle$ is the average of the additional force (with respect to the standard AdResS) one would have if the intermolecular force is calculated from Eq.\ref{hinter} as it happens in the Hamiltonian approach. This provided some physical arguments for the equality with $\Delta{\vect F}_\thf$ (besides the numerical evidence employed by us in first instance). In fact, if one writes down 
the average total force acting in the Hamiltonian approach and that in the standard AdResS, the difference between them is: ${\vect F}_\thf-{\vect F}_\thf^\hadress+ \langle w\nabla w(U^{\AT}-U^{\CG}) \rangle$. Since $\Delta{\vect F}_\thf=\langle w\nabla w(U^{\AT}-U^{\CG})\rangle$, this implies that the total average force acting in the two approaches is the same, which further implies that the two approaches are essentially equivalent from a numerical point of view as one would expect since they have the same equilibrium.
However, this procedure is based mainly on practical considerations and numerical experiments, thus at this stage it must be considered by the reader as a preliminary procedure although the essential ingredients seems to be already well defined.
Instead, an elegant procedure for adaptive simulation based on the interpolation of potentials and rigorous thermodynamics considerations has been recently presented by Potestio {\it et al.} \cite{H-adress}. Finally, as already stated before in this work, the approach based on the Hamiltonian shall not be considered a proper Hamiltonian approach to adaptive resolution, but only an auxiliary tool to determine in simple way a specific quantity of interest. In fact, if this is considered a proper Hamiltonian approach, one would have the physical inconsistency underlined in Ref.\cite{presolo}, that is the inevitable fact that the total potential (i.e. interpolated potential plus the potential corresponding to the thermodynamic force) in either the coarse-grained region or the atomistic region is not that of the original system (as it should be in a proper adaptive Hamiltonian procedure). However, in our view, the most important limitation is that, as shown in the current work, in this case the numerical accuracy can be assured only at the first order and there would not be a systematic way to improve it. This is the difference with the force interpolation-based method proposed here, where corrective forces in the transition region impose numerical correctness of the probability distribution of the system at higher orders. At this point, according to the procedure presented above, we have all the ingredients to calculate the excess chemical potential using AdResS and compare it with the same quantity calculated with the IPM, this is done in the next section.}

\subsection{Calculation of $\mu_\exc$ and numerical experiments}
\redc{For the difference of the total chemical potential between coarse-grained and atomistic region we have:
\begin{equation}
\mu^\AT-\mu^\CG=\int {\vect F}_\thf \dd dx+\omega_\thermo^\ext+\omega_\dof.
\label{mu}
\end{equation}
{The excess chemical potential is defined by removing the kinetic contribution from the total chemical potential, i.e. $\mu_\exc=\mu-\mu_\kin$}, ($\mu_\kin$ is the kinetic part of the
chemical potential, which can be calculated analytically).
It follows that $\mu_\exc^\AT-\mu_\exc^\CG=\int {\vect F}_\thf dx+\omega_\thermo^\ext + \omega_\dof - \Delta \mu_\kin = \int [{\vect F}_\thf+ \langle w\nabla w(U^{AT}-U^{CG})\rangle] dx + \omega_\dof - \Delta \mu_\kin$. Next, by knowing $\mu_\exc^\CG$, which can be calculated a rather low cost with IPM, one obtains $\mu_\exc^\AT$, whose calculation is instead very expensive with IPM.\\

We have carried out several numerical tests interfacing spherical systems with different representation; we have found always a very satisfactory agreement between the chemical potential calculated with AdResS and that calculated with IPM. More importantly, we have applied this idea to liquid water and found a value of $-22.8\, \textsf{kJ/mol}$ with AdresS and a value of $-24.6\, \textsf{kJ/mol}$ with IPM (still not fully converged after about 400\,\textsf{ns}), which must be compared with the value from literature of $-23.5\, \textsf{kJ/mol}$~\cite{flo}.
The difference between the value from our approach and that from IPM is about $8\%$, which is not significant, but the computational costs of the IMP is much larger than that based on the use of AdResS.} 
\end{document}